\documentclass{aa}
\usepackage{graphicx}
\usepackage{txfonts}
\usepackage{hyperref}
\graphicspath{{Images/}{../Images/}}
\usepackage{xcolor}
\usepackage[normalem]{ulem}

\begin{document}

    \title{Evolution of Jupiter and Saturn with helium rain} %evolutions with theoretical and experimental hydrogen-helium phase diagrams}
    %\title{Jupiter and Saturn's evolution with different hydrogen-helium phase diagrams}
    %\title{Theoretical and experimental hydrogen-helium phase diagrams applied to Jupiter and Saturn's evolutions}

    \author{S. Howard \inst{1}
    \and S. Müller \inst{1}
    \and R. Helled \inst{1}
          }

    \institute{Institut für Astrophysik, Universität Zürich, Winterthurerstr. 190, CH8057 Zurich, Switzerland,\\
              \email{saburo.howard@uzh.ch}
        %\and
        %     Institute for Computational Science, Center for Theoretical Astrophysics \& Cosmology, University of Zurich, Winterthurerstr. 190, CH8057 Zurich, Switzerland,
        }
    \date{}
    
% \abstract{}{}{}{}{} 
% 5 {} token are mandatory
 
  \abstract
  {The phase separation between hydrogen and helium at high pressures and temperatures leads to the rainout of helium in the deep interiors of Jupiter and Saturn. This process, also known as "helium rain," affects their long-term evolution. Modeling the evolution and internal structure of Jupiter and Saturn (and giant exoplanets) relies on the phase diagram of hydrogen and helium. In this work, we simulated the evolution of Jupiter and Saturn with helium rain by applying {different} phase diagrams of hydrogen and helium {and we searched for models that reproduce the measured atmospheric helium abundance in the present day}. %We wish to infer the internal composition and structure of both planets and to refine our understanding of the phase diagram. 
  We find that a consistency between Jupiter's evolution and the Galileo measurement of its atmospheric helium abundance can only be achieved if a shift in temperature is applied to the existing phase diagrams  (-1250~K, +350~K or -3850~K depending on the applied phase diagram). Next, we used the shifted phase diagrams to model Saturn's evolution and we found consistent solutions for both planets. We confirm that de-mixing in Jupiter is modest, whereas in Saturn, the process of helium rain is significant. We find that Saturn has a large helium gradient and a helium ocean. Saturn's atmospheric helium mass fraction is estimated to be between 0.13 and 0.16. We also investigated how the applied hydrogen-helium equation of state and the atmospheric model affect the planetary evolution, finding that the predicted cooling times can change by several hundred million years. 
  Constraining the level of super-adiabaticity in the helium gradient formed in Jupiter and Saturn remains challenging and should be investigated in detail in future research. 
  %\sout{Overall, we show that including helium rain, from adjusted phase diagrams of hydrogen and helium, into the evolutionary calculations of Jupiter and Saturn can explain the observed luminosities of these planets. We suggest that} 
 We conclude that further explorations of the immiscibility between hydrogen and helium are valuable as this knowledge directly affects the evolution and current structure of Jupiter and Saturn. Finally, we argue that measuring Saturn's atmospheric helium content is crucial for constraining Saturn's evolution as well as the hydrogen-helium phase diagram.
   }
  % context heading (optional)
  % {} leave it empty if necessary 
  % aims heading (mandatory)
   %We use different state-of-the-art theoretical and experimental phase diagrams of hydrogen and helium and infer the required shifts... 
   %\qsh{given that both planets share fundamental properties of hydrogen and helium}. 
   %This constrains thephase diagram and also constraint...
  % methods heading (mandatory)
  % results heading (mandatory)
  % conclusions heading (optional), leave it empty if necessary 
   

   \maketitle
%
%-------------------------------------------------------------------
%-------------------------------------------------------------------

\section{Introduction}
%-------------------------------------------------------------------
%-------------------------------------------------------------------
\label{section:1}

Ground-based infrared observations from \citet{low1966} revealed that Jupiter and Saturn emit more energy than they receive from the Sun. Their observed luminosities cannot be explained solely by the continued cooling and contraction of these planets and it was suggested that both planets possess an additional internal heat source. The origin of this heat source was proposed to be the differentiation of helium, due to a phase separation between hydrogen and helium \citep[H-He; e.g.,][]{smoluchowski1967,salpeter1973}. As helium becomes immiscible with metallic hydrogen \citep{stevenson1975,stevenson1977a} present in the interior of giant planets, helium droplets can form and sink (known as helium rain), providing an additional source of thermal energy. Homogeneous evolution models which do not include change of composition with time, have consistently found that Jupiter's cooling time is roughly in line with the age of the Solar System 
\citep{hubbard1969}. With the age of the Solar System being $4.56 \pm 0.10\,$Gyr \citep{connely2012}, the inferred cooling times of 4.2 to 5.3~Gyr for Jupiter \citep{graboske1975,hubbard1977,saumon1992,guillot1995_EvRad} indicate that helium de-mixing may not have  started at all or started only recently \citep{stevenson1977b}. On the contrary, similar models for Saturn predict significantly shorter cooling times, between 2 and 2.7~Gyr \citep{pollack1977,saumon1992,fortneynettelmann2010}, suggesting a strong excess of luminosity. This, in turn, suggests that in Saturn helium de-mixing is significant.

Nevertheless, the excess luminosity observed for Jupiter and especially Saturn originally motivated the idea of helium de-mixing; however, this does not serve as undeniable evidence for helium rain occurring in their interiors \citep{stevenson2020}. For example, the presence of a heavy-element gradient, already suggested by \citet{stevenson1985} and supported by recent interior models constrained by Juno and Cassini measurements \citep{wahl2017,debras2019,mankovich2021,miguel2022,militzer2022,howard2023_interior} could also significantly affect the thermal evolution of giant planets. In this scenario, heat would initially be trapped in the deep interior with the composition gradient acting as a thermal barrier. It could then be released later, which would explain such an excess in luminosity \citep{leconte2013,vazan2016}.

The strongest indication of helium rain in Jupiter is still the in situ measurements of the Galileo probe \citep{vonzahn1998}, which showed a slight depletion of helium ($Y_{\rm atm}=0.238 \pm 0.005$) compared to the protosolar value ($Y_{\rm proto}=0.27$) \citep{asplund2021} and also neon depletion, which could be sequestered in the falling helium droplets \citep{wilson2010}. Saturn's atmospheric helium content remains uncertain and needs to be measured by a dedicated atmospheric entry probe \citep{mousis2016,fortney2023}.

Meanwhile, progress in improving our understanding of the H-He phase diagram has enabled the calculation of inhomogeneous evolution models, thereby accounting for helium rain. Such calculations from \citet{hubbard1999} and \citet{fortney2003} based on early H-He phase diagrams \citep{HDW1985,pfaffenzeller1995} confirmed that helium rain could explain Saturn's measured luminosity. Extensive ab initio calculations were then conducted to model the behaviour of H-He mixtures at high pressures. \citet{lorenzen2009,lorenzen2011} performed molecular dynamic simulations based on density functional theory (DFT-MD) on a large range of H-He mixtures, using the Perdew-Burke-Ernzerhof (PBE) exchange-correlation (XC) functional and the ideal entropy of mixing approximation. The resulting phase diagram was employed in evolution calculations of Jupiter  \citep[e.g.,][]{nettelmann2015,mankovich2016} and  Saturn \citep[e.g.,][]{pustow2016}. The ideal entropy of mixing approximation was avoided by \citet{morales2009} and \citet{morales2013}, leading to lower de-mixing temperatures. \citet{schottler2018_prl} provided the latest theoretical phase diagram, using the van der Waals density functional (vDW-DF) which led to even lower de-mixing temperatures. Recently, \citet{mankovich2020} searched for  consistent solutions for Jupiter's and Saturn's evolution by slightly shifting the phase diagram in temperature calculated by \citet{schottler2018_prl}. This was only possible by assuming an enhanced Bond albedo for Saturn. Surprisingly, a laser-shock experiment \citep{brygoo2021} yielded much higher de-mixing temperatures than the theoretical predictions (see Fig.~\ref{figure:phase_diagrams}).

Understanding giant planets requires a good comprehension of the H-He phase diagram, yet it remains very uncertain given the important discrepancy between theoretical calculations and single experiments. In this study, we applied the existing phase diagrams to the evolution of Jupiter and Saturn. We then searched for a coherent solution between the evolution of Jupiter and Saturn since the same fundamental properties of hydrogen and helium are at play. We find that this can be achieved by a temperature shift of the existing phase diagrams and, as we discuss below, this could provide hints about the location and shape of the exact phase diagram. Such an analysis may also be able to constrain the internal structure of these planets.

%-------------------------------------------------------------------
%-------------------------------------------------------------------

%-------------------------------------------------------------------
%-------------------------------------------------------------------
\section{Methods}
\label{section:2}

\subsection{H-He phase diagrams}
\label{subsec:phase_diag}

\begin{figure}[h]
   \centering   \includegraphics[width=\hsize]{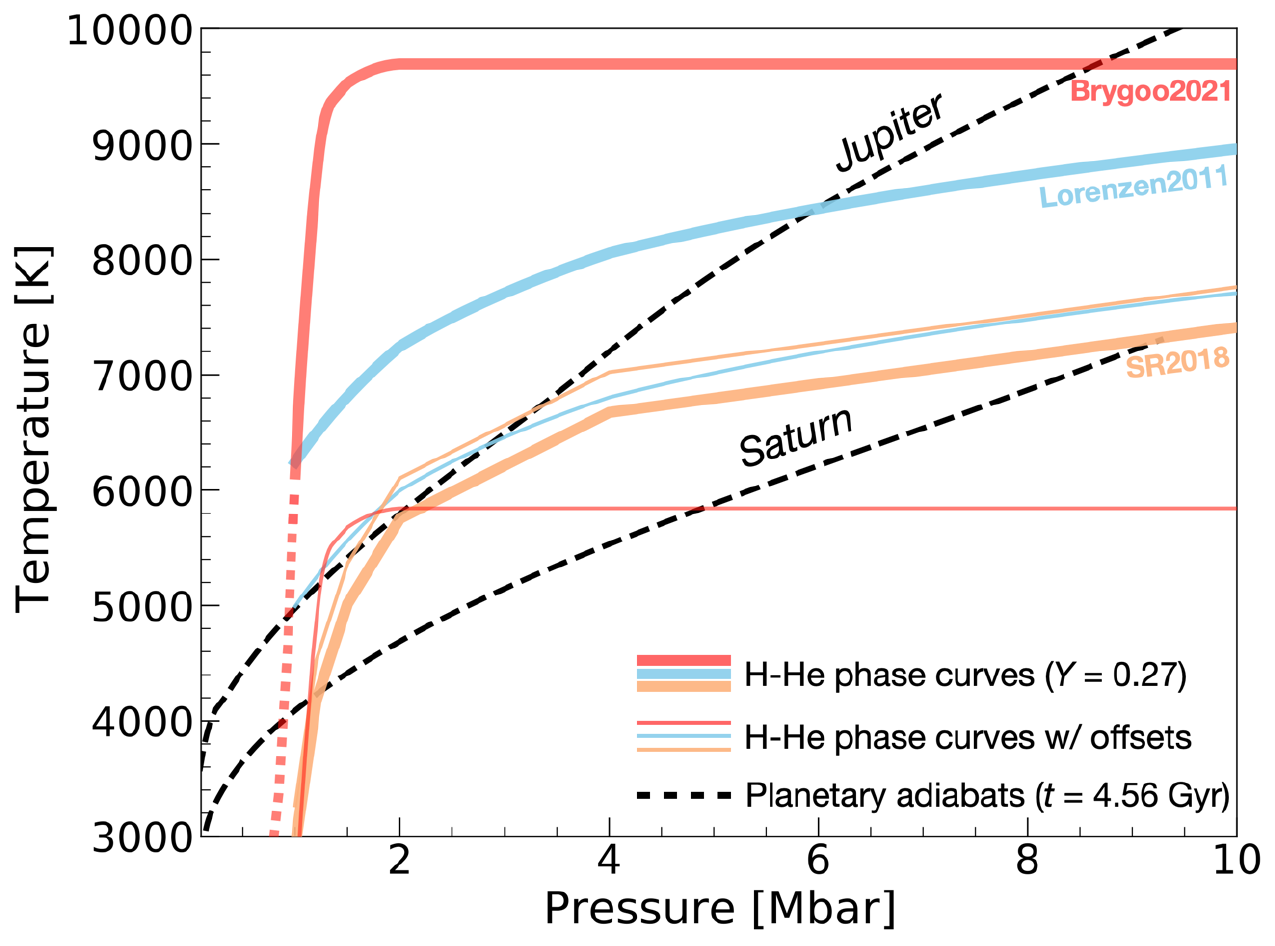}
      \caption{Comparison of immiscibility curves from H-He phase diagrams. The thick solid lines show the phase curves for a solar mixture from ab initio calculations of \citet{lorenzen2011} and \citet{schottler2018_prl} and from the laser-shock experiment of \citet{brygoo2021}. The thin solid lines show the same phase curves but with temperature offsets of -1250, +350, and -3850~K, respectively. These offsets correspond to the ones used in Sect.~\ref{subsubsec:jupiter}, which are required to fit Jupiter's evolution to Galileo's measurement of its atmospheric helium content. The dashed lines correspond to the adiabats of Jupiter and Saturn today (see Sect.~\ref{section:3} for details). As the region of de-mixing was considered only at pressures greater than 1~Mbar in our models, the region of the Brygoo2021 phase diagram below 1~Mbar was not used (shown with the dotted red line).}
         \label{figure:phase_diagrams}
\end{figure}

In this study, we used the theoretical phase diagrams of \citet{lorenzen2011} (hereafter, Lorenzen2011) and \citet{schottler2018_prl} (SR2018), as well as {a constructed} phase diagram {based on experimental data} \citet{brygoo2021} (Brygoo2021).
These three phase diagrams are shown (for a H-He mixture in protosolar proportions) in Fig.~\ref{figure:phase_diagrams}. 
These phase diagrams are then coupled to the evolution calculations of Jupiter and Saturn. This requires us to have access to the phase curves at several helium fractions, $x_{\rm He}$, and in various $P-T$ conditions. 
Therefore, we linearly interpolated the data from \citet{lorenzen2011} and \citet{schottler2018_prl} to obtain the de-mixing temperatures on a rectangular grid with 301 values of $x_{\rm He}$ between 0 and 1 and 301 values of $P$ between 1 and 24~Mbar. The \citet{brygoo2021} data provides only one curve at $x_{\rm He}=0.11$ ($Y=0.33$). We constructed a more complete phase diagram (on the same grid as Lorenzen2011 and SR2018) in the following way. We used the SR2018 phase diagram and estimate the temperature differences at each pressure value between the curve at $x_{\rm He}=0.11$ and the curves at all other $x_{\rm He}$ values. Next, we applied these temperature differences to the single curve from \citet{brygoo2021}. Figure~\ref{figure:phase_diagrams} also shows the phase curves of the Lorenzen2011, SR2018 and Brygoo2021 phase diagrams shifted by temperature offsets of $T_{\rm offset}=-1250$, +350 and -3850~K, respectively.

These offsets must be applied to infer a depletion of helium in the atmosphere of Jupiter that is consistent with the Galileo measurement by the evolution model (more details in Sect.~\ref{subsec:fiducial}). {The value of these offsets depend on the used H-He equation of state.} Given the high discrepancy between the three original phase diagrams employed here, it is appropriate to allow for temperature offsets when studying the evolution of Jupiter and Saturn \citep[e.g.,][]{nettelmann2015,mankovich2016,pustow2016,mankovich2020}. 

\subsection{Calculation of the helium profile}
\label{subsec:he_profile}

Applying H-He phase diagrams to the evolution of giant planets allows for calculating the helium distribution throughout their evolution. We follow the procedure of an earlier work, which is aptly described in, for instance, \citet{nettelmann2015} and \citet{mankovich2016} and also illustrated in Figure 2 of \citet{mankovich2020}. The main ideas are summarized below. 

Starting from an initial "hot and puffed" state, giant planets cool down and their pressure-temperature profile eventually intersects with a region of the phase diagram, which predicts de-mixing between hydrogen and helium (see Fig.~\ref{figure:phase_diagrams}). Helium droplets start to form in the intersecting region and then sink to deeper regions. Helium can then mix again in the deeper region if de-mixing is not expected. Such a mechanism is possible due to the hierarchy on different timescales \citep{stevenson1977a,mankovich2016}, namely: the diffusion timescale of the helium droplets ($\sim~$$10^{-1}~$s), timescale of convective mixing ($\sim~$yr), and the timescale corresponding to the long-term thermal planet evolution ($\sim~$Gyr). The region of the planet where phase separation occurs is depleted in helium until it achieves exact saturation, namely, this means that it has reached a new equilibrium abundance where no further separation occurs. As the planet continues to cool down, there might be a new equilibrium abundance and, thus, helium can be depleted further. The region above phase separation will also be depleted as it supplies helium to the region undergoing phase separation when convective mixing occurs. The inner part of the planet is progressively enriched in helium.

%The region of the planet where phase separation occurs is depleted in helium until it has lost enough helium so that phase separation does not occur anymore.

In the following, we describe how we proceeded with our evolutionary models. At every time step and in each layer of our model, the temperature $T(P,x_{\rm He})$ is compared to the predicted de-mixing temperature, $T_{\rm de-mix}(P,x_{\rm He})$. As we also allowed for a temperature offset on the phase diagram, if $T(P,x_{\rm He})<T_{\rm de-mix}(P,x_{\rm He})+T_{\rm offset}$, de-mixing occurs. We calculated the helium concentration for which both terms of the inequality are equal to one another and we find the new helium abundance, $x'_{\rm He}$, after de-mixing (for exact saturation). The region undergoing phase separation yields one minimum value, $x'_{\rm He, min}$, at a pressure denoted as $P_0$. As our evolution timestep ($\sim~$10-100 Myr) is larger than the convective mixing timescale, we assign $x'_{\rm He, min}$ to every layer with $P \leq P_0$. The shape of the phase diagram then naturally builds a helium gradient in the phase separating region, where $P>P_0$. The helium lost by de-mixing is redistributed uniformly in the region below the location where phase separation occurs. If de-mixing occurs in the very deep interior near the planetary center or if enough helium sinks, a helium ocean forms (i.e., a {region dominated by helium}). 

\subsection{Evolution model setup}
\label{subsec:model_setup}

We calculated the evolution of Jupiter and Saturn using CEPAM \citep{guillot1995_cepam} {with our key objective being to find evolution models that reproduce the measured atmospheric helium abundance at present day}. We solved the classical structure equations \citep[see, e.g.,][]{guillot2005,helled2018,miguel2023,RavitSaburo} as follows:
\begin{equation}
    %\textrm{Hydrostatic equilibrium: }
    \frac{\partial P}{\partial r} = -\rho g,
    \label{eq:basic_hydroeq}
\end{equation}
\begin{equation}
    %\textrm{Mass conservation: }
    \frac{\partial m}{\partial r} = 4 \pi r^2 \rho,
\end{equation}
\begin{equation}
    %\textrm{Energy transport: }
    \frac{\partial T}{\partial r} = \frac{\partial P}{\partial r} \frac{T}{P} \nabla_T,
\end{equation}
\begin{equation}
    %\textrm{Energy conservation: }
    \frac{\partial L}{\partial r} = -4 \pi r^2 \rho  T \frac{\partial S}{\partial t},
    \label{eq:energy_conservation}
\end{equation}
where $P$ is the pressure, $r$ is the radius, $\rho$ is the density, $g$ is the gravitational acceleration, $m$ is the mass, $T$ is the temperature, $\nabla_T=\frac{\textrm{dln}T}{\textrm{dln}P}$ is the temperature gradient, $L$ is the intrinsic luminosity, and $S$ is the specific entropy (noting that we neglected the rotation). Equation~\ref{eq:energy_conservation} is particularly relevant in the context of evolution calculations. The change in composition over time due to helium differentiation (Sect.~\ref{subsec:he_profile}) is accounted for in the entropy time derivative term.

We describe here the setup of our baseline models (presented in Sect.~\ref{subsec:fiducial}). We modeled the planets based on a central core, made of 50\% rocks and 50\% ices, surrounded by a homogeneous H-He envelope that follows protosolar proportions. All the heavy elements are concentrated in the core and the core mass should be interpreted as a total heavy-element mass (see \citet{mankovich2016}). The assumption of a pure H-He envelope is consistent with the existing phase diagrams calculations that correspond to pure H-He mixtures with no heavy elements (more discussion in Sect.~\ref{section:4}). 

Our baseline models use a core mass of $30~M_{\oplus}$ for Jupiter and $20~M_{\oplus}$ for Saturn, which correspond to the total heavy-element mass inferred from recent interior models based on gravity and seismology data \citep{mankovich2021,miguel2022,howard2023_interior}. Different core masses have been tested and results are presented in Sect.~\ref{subsubsec:core_mass}. 
The assumed structure of our models is simplified; the need to consider more realistic distributions of heavy elements is discussed in Sect.~\ref{section:4}.
The helium gradient formed by helium de-mixing is kept adiabatic in our models. 
However, a super-adiabatic temperature gradient may remain stable against convection in the presence of a composition gradient \citep{ledoux1947}. Such a possibility is discussed in Sect.~\ref{subsubsec:superadiabaticity}.
%As we discuss in Sec.~\ref{section:4}, more realistic distributions of heavy elements would be important to consider in future but are beyond the scope of this study.} 

Prior evolution models of Jupiter and Saturn with helium rain used outdated equations of state (EOSs), which do not properly account for H-He interactions \citep[e.g.,][]{nettelmann2015,mankovich2016,pustow2016}. \citet{mankovich2020} used the MH13 EOS \citep{militzer2013}, which includes non-ideal mixing effects; however for a constant composition ($Y=0.245$). Our baseline models use the CMS19+HG23 EOS, which combines pure hydrogen and helium tables from \citet{chabrier2019} with non-ideal mixing effects from \citet{howard2023}. The specificity of this EOS is that non-ideal mixing effects depend on composition. We recall that the corrections on density or entropy are defined as $\Delta V (X,Y)=XYV_{\rm mix}$ and $\Delta S (X,Y)=XYS_{\rm mix}$, where $X$ and $Y$ are the mass fractions of hydrogen and helium and $V_{\rm mix}$ and $S_{\rm mix}$ are tabulated quantities from \citet{howard2023}.

Furthermore, we used the atmospheric models of \citet{fortney2011}. As in \citet{mankovich2020}, we recalculated the Jupiter table to make it consistent with the Cassini Bond albedo measurement \citep{li2018}. We also accounted for the evolution of the Sun's luminosity by interpolating linearly from $0.7~L_{\odot}$ at 0~Gyr to $1.0~L_{\odot}$ at 4.56~Gyr. The effects of the EOS and of the atmospheric model on evolution calculations are presented in Sects.~\ref{subsubsec:eos} and~\ref{subsubsec:atm_model}, respectively. 
\par
\begin{figure}[h]
   \centering   
   \includegraphics[width=\hsize]{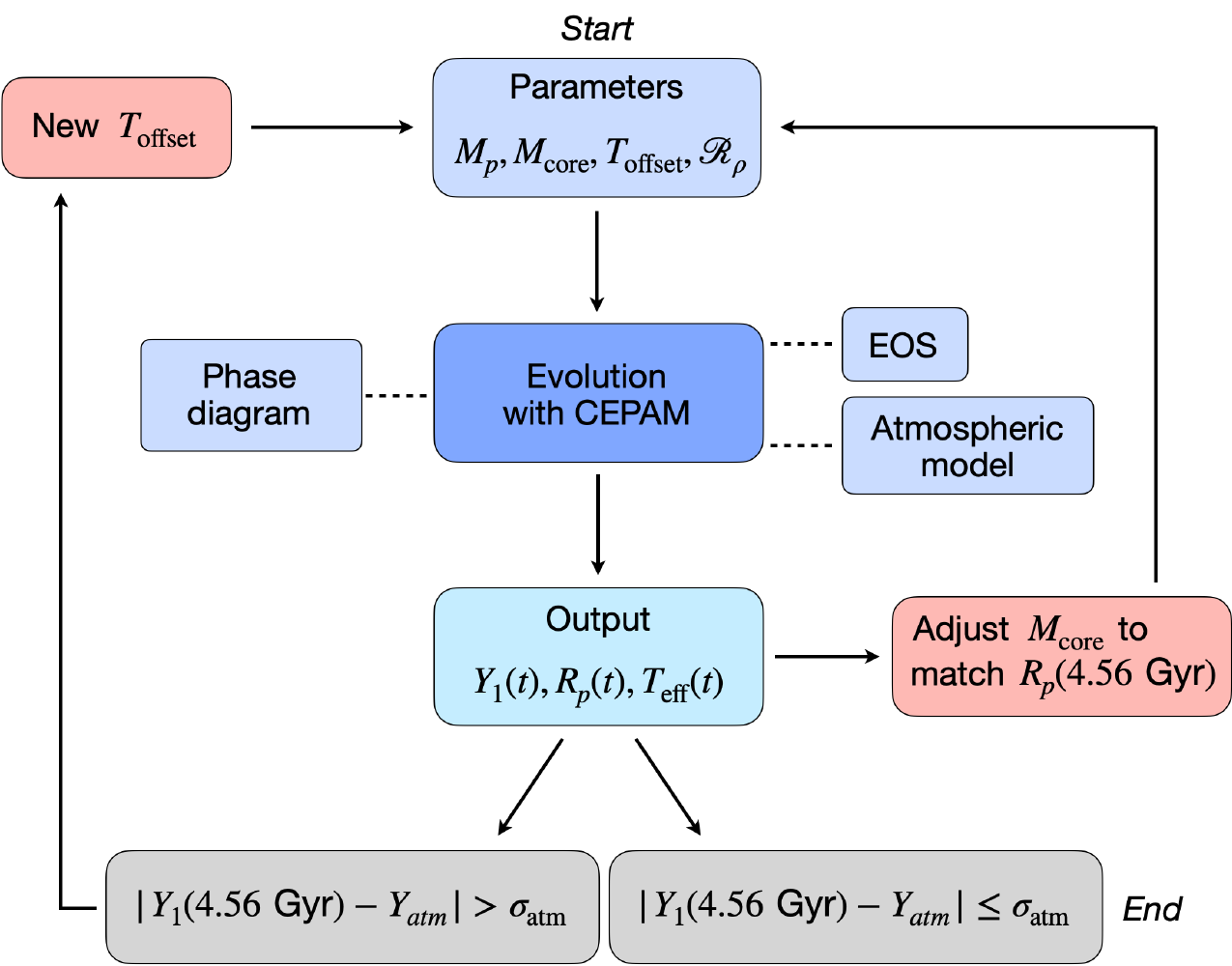}
      \caption{{Flowchart of our evolution calculations. The input parameters $M_{p},M_{\rm core},T_{\rm offset},  \mathcal{R}_\rho$ are the planetary mass, the core mass, the temperature offset, and the density ratio, respectively (see Sect.~\ref{subsubsec:superadiabaticity}). CEPAM calculates the atmospheric helium mass fraction, the planetary radius, and the effective temperature as a function of time. If the inferred atmospheric helium mass fraction at the present time (4.56~Gyr) does not fall within the uncertainty range of the measured value by Galileo, we updated $T_{\rm offset}$ and repeated the procedure.}}
         \label{figure:flowchart}
\end{figure}

{Figure~\ref{figure:flowchart} summarizes the procedure we followed.  Once the input parameters of the model (planetary mass, core mass, and temperature offset) were set, we calculated the planetary evolution, using a chosen EOS, atmospheric model, and phase diagram. At the end of the evolution, we compared the calculated atmospheric helium mass fraction at $t=4.56~$Gyr (the current-age of the planets)   to the measured value ($Y_{\rm atm}=0.238 \pm 0.005$ from Galileo \citep{vonzahn1998}). If it is not within the uncertainty bounds, we modified the temperature offset and run again the evolution calculation until the measured value is obtained. The radius and the effective temperature as a function of time were also calculated, along with the output parameters. As we present in Sect.~\ref{subsubsec:core_mass}, the radius can be adjusted by changing the core mass (or atmospheric model). In Appendix~\ref{appendix:t1bar}, we show the results of the evolution models when we used the constraint on the 1 bar temperature instead of the planetary age.} The measured values of effective temperature ($T_{\rm eff}$), radius ($R$), age, and atmospheric abundance of helium ($Y_1$) are similar to those in \citet{mankovich2020} and we reproduce their table here (see Table~\ref{tab:param}). 

\begin{table}[h]
\centering
\caption{Jupiter and Saturn measured parameters, adapted from \citet{mankovich2020}. See references therein.}
\begin{tabular}{l c c}
\hline
\hline
 &  Jupiter & Saturn \\
\hline
$T_{\rm eff}$ [K]  & $125.57 \pm 0.07$ & $96.67 \pm 0.17$ \\ 
$R$ [km] & $69,911 \pm 6$ & $58,232 \pm 6$ \\ 
Age [Gyr] & $4.56 \pm 0.1$ & $4.56 \pm 0.1$ \\
$Y_1/(X_1+Y_1)$ & $0.238 \pm 0.005$ & - - \\
\hline
\end{tabular}
\label{tab:param}
\begin{flushleft}
\end{flushleft}
\end{table}

%Initial radius. No rotation.

%We finally stress that despite certain limitations in our study that could be refined in future research, we think the relative comparison of our results, grounded in the same framework, can provide meaningful insights. We did not conduct extensive model calculations as done by \citet{mankovich2020} with a Bayesian analysis. 

%-------------------------------------------------------------------
%-------------------------------------------------------------------
\section{Results}
  \label{section:3}

\subsection{Baseline models}
    \label{subsec:fiducial}

\subsubsection{Jupiter}
\label{subsubsec:jupiter}

\begin{figure*}
\centering
   \includegraphics[width=17cm]{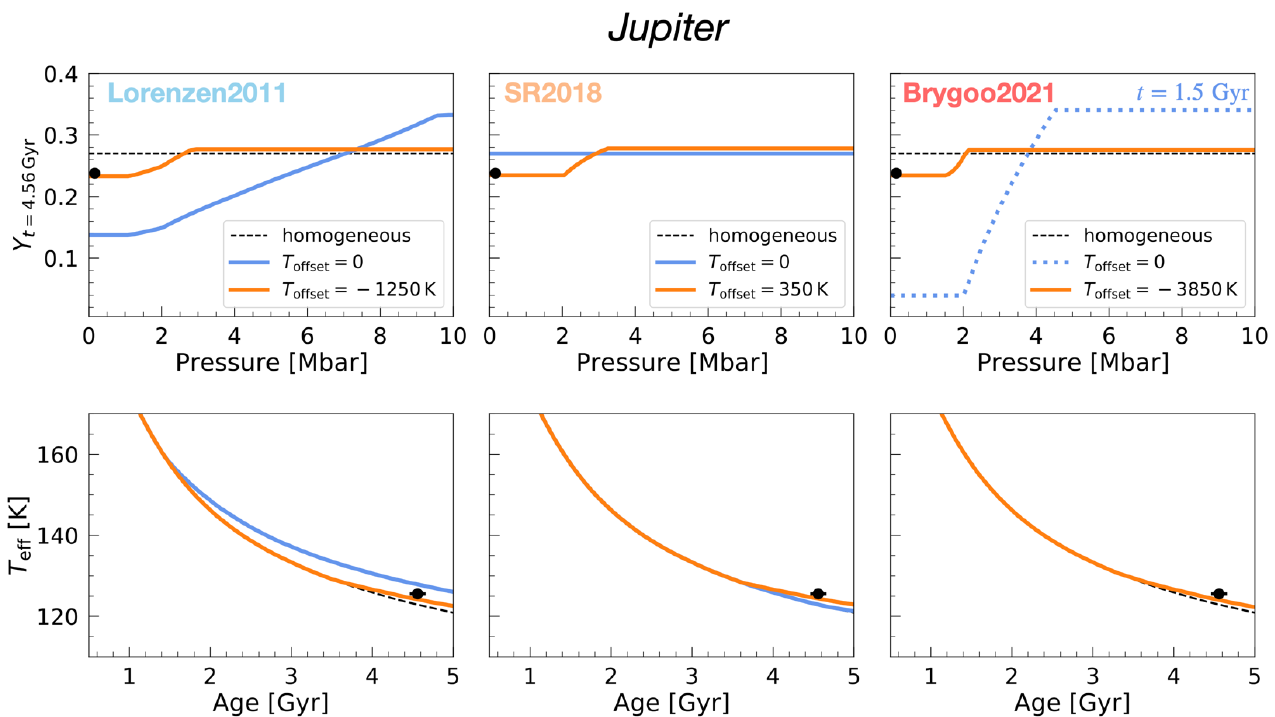}
     \caption{Evolutionary calculations of Jupiter. \textit{Top panels:} Present-day helium mass fraction as a function of pressure. The pressure range corresponds to the relevant region where de-mixing takes place. \textit{Bottom panels:}  Effective temperature as a function of age. Columns (\textit{from left to right}) correspond to results where we applied the Lorenzen2011, SR2018 and Brygoo2021 phase diagrams, respectively. For comparison, black dashed lines show homogeneous evolution calculations, namely with no de-mixing. Blue solid lines show evolution calculations coupled with the original phase diagrams, namely with no temperature offset. For Brygoo2021, the blue line is dotted and shows the distribution at $t=1.5~$Gyr. The orange solid lines show results where a temperature offset, $T_{\rm offset}$, was applied to the corresponding phase diagram in order to fit the abundance of helium measured by Galileo (represented by the black dot on top panels). The black dot in the bottom panels corresponds to  the measured effective temperature and the errorbar refers to the uncertainty in age.}
     \label{figure:Jupiter_HG23}
\end{figure*}

We first ran the evolution models for Jupiter, coupled with the original Lorenzen2011, SR2018, and Brygoo2021 phase diagrams. We identified the required temperature offsets and applied them to the original phase diagrams, so that evolution calculations yield an atmospheric helium abundance that is in agreement with the Galileo measurement ($Y_{\rm atm}=0.238 \pm 0.005$). The results are shown in Fig.~\ref{figure:Jupiter_HG23}. The original Lorenzen2011 phase diagram leads to too much de-mixing at $t=4.56~$Gyr, while the SR2018 indicates that de-mixing has not started yet. These results are consistent with previous models from \citet{nettelmann2015}, \citet{mankovich2016} and \citet{mankovich2020}. Interestingly, the Brygoo2021 phase diagram yields $Y_1=0.04$ after only 1.5~Gyr. In this case, de-mixing starts already 500~Myr after Jupiter's formation and we would expect an overly { large} depletion of helium in Jupiter's outer envelope in the present era. The Galileo probe measured only a slight depletion in Jupiter's atmosphere compared to protosolar. We therefore conclude that the original Brygoo2021 phase diagram appears to be very inconsistent with Jupiter's evolution.

By shifting the Lorenzen2011, SR2018 and Brygoo2021 phase diagrams with adequate offsets (-1250, +350 and -3850~K, respectively), we can fit Jupiter's evolution to Galileo's measurement. {These offsets are determined with an accuracy of approximately $\pm 50~$K, as variations within this range ensure that the inferred atmospheric helium abundance stays within the uncertainty bounds of Galileo's measurement.} {We stress again that the values of these offsets hold only in conjunction with the chosen setup and notably the applied EOSs. In comparison, \citet{nettelmann2015}, using the Lorenzen2011 phase diagram, only required $T_{\rm offset}=-200$~K to match Jupiter's observed atmospheric helium abundance. However, they used the older SCvH EOS \citep{saumon1995}, which leads to a warmer planetary adiabat by about 1000~K. Similarly, \citet{mankovich2016} inferred an offset of the same order of magnitude. Using the SR2018 phase diagram and the MH13 EOS, \citet{mankovich2020} required $T_{\rm offset}=539 \pm 23~$K. The H-He EOS is crucial as it dictates the interior temperature; hence, it affects the intersection with the phase diagram.} 

Here, we only intended to constrain the phase diagrams (using $T_{\rm offset}$) by fitting the outer envelope helium abundance yielded by Jupiter's evolution at $t=4.56~$Gyr to Galileo's value. Still, we found an effective temperature for Jupiter that is relatively close to the measured value. {We found $T_{\rm eff}$ lower than the measured value by up to 1.5~K depending on the applied phase diagram. Homogeneous models that do not account for helium rain predict $T_{\rm eff}$ lower by 2.7~K. Such homogeneous models predict a cooling time of $\sim~$4.0~Gyr.}
For the three shifted phase diagrams, we find a cooling time slightly lower than the age of the planet, by only about $\sim 0.2$~Gyr, which is smaller than the uncertainty due to some modeling aspects (Sect.~\ref{subsection:effects}). The onset of helium differentiation occurs at ages between 3.5 and 3.7~Gyr, {in agreement with \citet{nettelmann2015}, at 3.7~Gyr, and \citet{mankovich2020}, 3.6-4.0~Gyr}. The helium gradient in Jupiter remains small (from $Y_1 \sim 0.238$ to $Y_2 \sim 0.28$). It covers a narrow range of pressures and a small portion of the planet (from about 1 to 3~Mbar), {in line with \citet{nettelmann2015} and \citet{mankovich2016}}. 
It should be noted that not only the location of the phase diagram is important but also its shape. The Brygoo2021 phase diagram exhibits a plateau at $P>2~$Mbar and yields a steeper helium gradient compared to the two other phase diagrams.

\begin{figure*}
\centering
   \includegraphics[width=17cm]{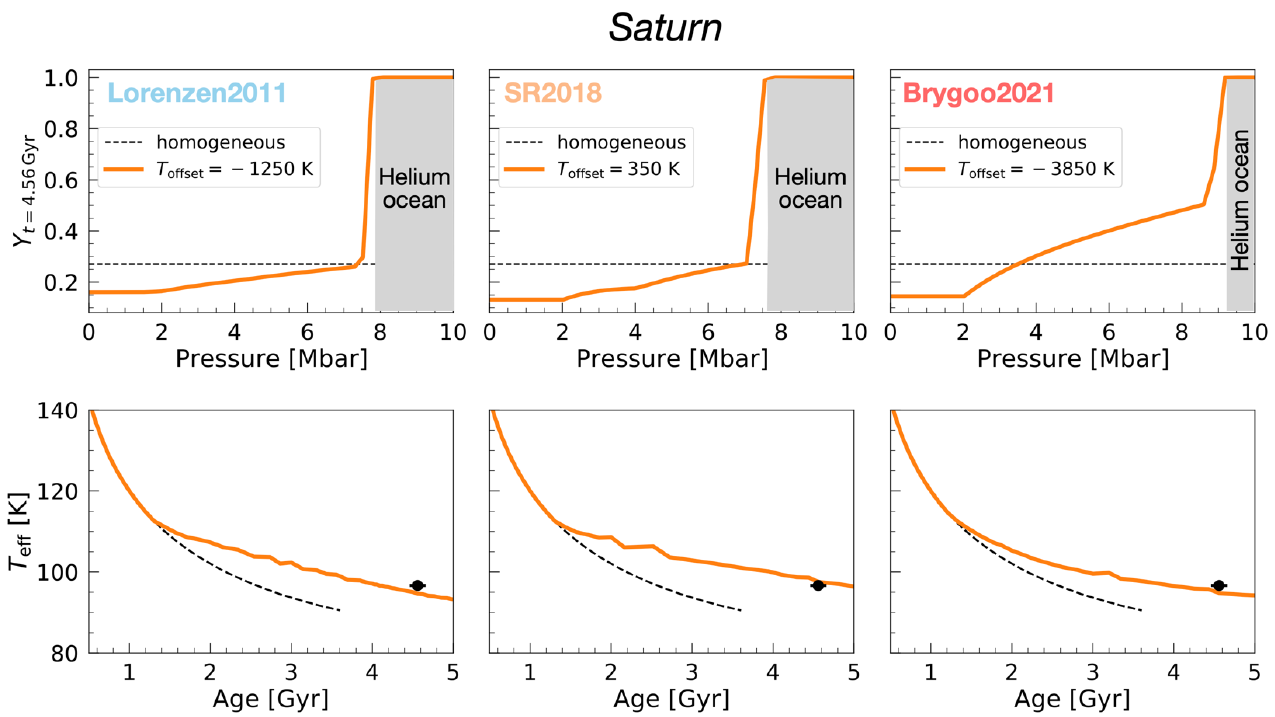}
     \caption{Evolutionary calculations of Saturn. Description is same as for Fig.~\ref{figure:Jupiter_HG23}. The shaded areas correspond to a "helium ocean," a region above the core where the mass fraction of helium reaches unity. %Various shades of grey emphasize the varying sizes of this helium ocean between calculations with different phase diagrams. 
     The homogeneous evolution calculations represented by black dashed lines stop because the atmospheric models from \citet{fortney2011} do not cover lower effective temperatures.}
     \label{figure:Saturn_HG23}
\end{figure*}

\subsubsection{Saturn}
\label{subsubsec:saturn}

%We expect hydrogen and helium to behave similarly in Jupiter and Saturn. 
{The same fundamental properties of hydrogen and helium are at play in Jupiter and Saturn.}
Therefore, to model the evolution of Saturn, we then applied the constrained phase diagrams, with the offsets required by Jupiter's evolution in line with the Galileo measurement. The results are shown in Fig.~\ref{figure:Saturn_HG23}.

As expected, because Saturn's interior is colder than Jupiter's, helium de-mixing is more pronounced. We found a depleted atmosphere (and hence outer envelope) with $Y_1$ values of 0.16, 0.13, and 0.14 for the {shifted} Lorenzen2011, SR2018,  and Brygoo2021 phase diagrams, respectively. 
These values are consistent with the recent estimates from Cassini, namely, with the lower bound of \citet{koskinen2018} (0.16--0.22) and the upper bound of \citet{achterberg2020} (0.075--0.13). {The models from \citet{nettelmann2015} yielded similar $Y_1$ values (0.12-0.15) by shifting the Lorenzen2011 phase diagram and using the SCvH EOS while \citet{mankovich2020} found values below 0.10 by shifting the SR2018 phase diagram and using the MH13 EOS (see Sect.~\ref{subsubsec:eos}).} 

Right above the heavy-element core, a region with $Y=1$ (referred to as the helium ocean) was formed, due to helium sinking and accumulating there. {For simplicity, we assume $Y=1$ in this helium ocean but He-rich phases of $Y=0.8-0.9$ are predicted by the Lorenzen2011 and SR2018 phase diagrams and the presence of heavy elements, not accounted here, would also suggest a region, where $Y$ is not exactly 1. Models from \citet{pustow2016} and \citet{mankovich2020} also harbour a helium ocean, where the mass fraction of helium is about 0.9 and 0.95, respectively.} The extent of this helium ocean depends on the chosen phase diagram, but in all cases, it is confined within the range of 20\% -- 30\% of Saturn's total mass. A significant helium gradient, from $Y_1 \sim 0.1$ to $Y_2 \sim 1$, covers a wide range of pressures (from about 1 to 9~Mbar) and a large part of the planet (30\% -- 70\% by mass). {The width of the helium gradient is within the value found by \citet{mankovich2020}, namely:\ 2-14~Mbar. However, it also depends on the presence and the size of a pure heavy-element core (see Sect.~\ref{subsubsec:core_mass}).} As for Jupiter, the steepness of the helium profile depends on the shape of the phase diagram, with Brygoo2021 yielding a different helium distribution than the Lorenzen2011 and SR2018 ones that are more comparable. 

The obtained effective temperatures are in quite good agreement with the measurements. {We found $T_{\rm eff}$ values lower than the measured value by 2.0 and 1.9~K for the shifted Lorenzen2011 and Brygoo2021 phase diagrams, respectively. The shifted SR2018 phase diagram leads to $T_{\rm eff}$ values that are larger by 0.9~K. Homogeneous models predict $T_{\rm eff}$ lower by about 10~K and therefore a cooling time that is off by about 2~Gyr (as mentioned in Sect.~\ref{section:1}).} The shifted Lorenzen2011 and Brygoo2011 phase diagrams lead to a cooling time slightly lower than the age of the planet, by 0.5~Gyr at most. On the contrary, the shifted SR2018 phase diagram yields a cooling time slightly larger, by about $\sim 0.1$~Gyr. Consequently, the goal of finding a consistent solution for the evolution of both Jupiter and Saturn by applying the same H-He phase diagram is nearly attained. We found the onset of helium differentiation at about 1.3~Gyr quite independently of the applied phase diagram. {This is in agreement with the predicted onset at 1-2~Gyr from \citet{pustow2016} and at 1.5~Gyr according to \citet{mankovich2020}.}

\begin{figure}[h]
   \centering
   \includegraphics[width=\hsize]{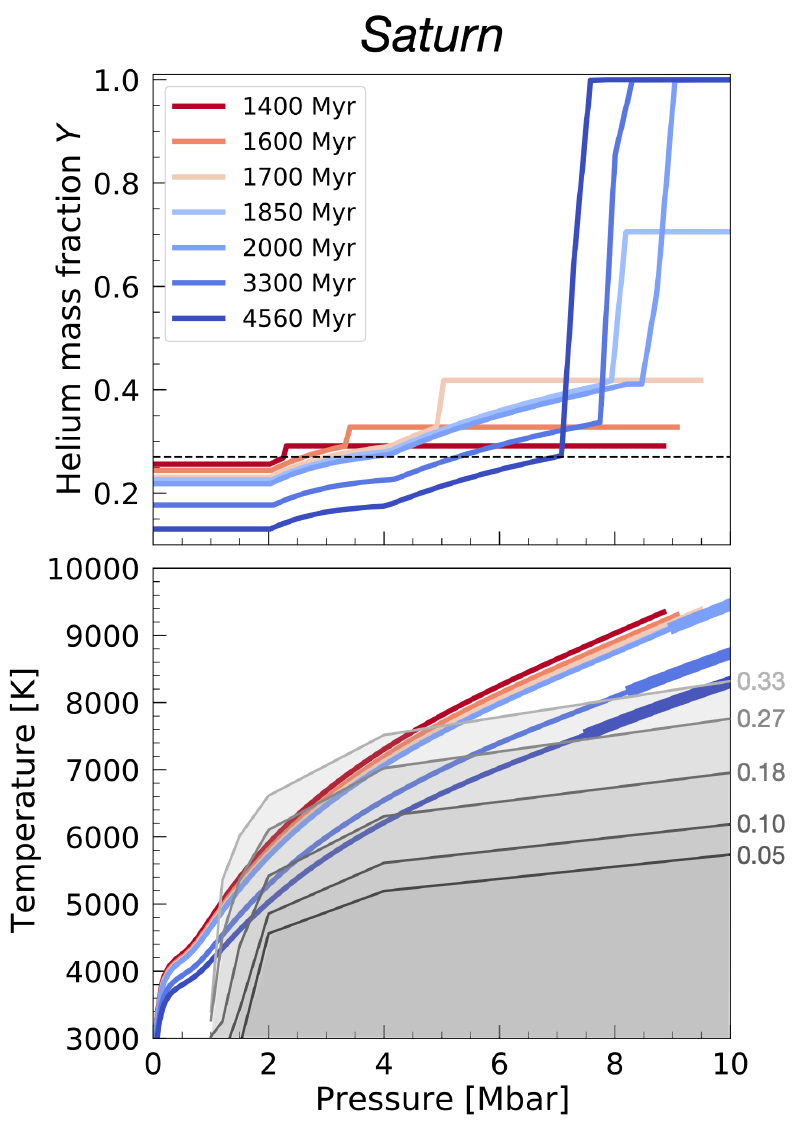}
      \caption{Growth of the helium ocean in Saturn. The calculations come from the results of Saturn's evolution using the SR2018 phase diagram with $T_{\rm offset}=350~$K (Sec~\ref{subsubsec:saturn}). \textit{Top panel:} Helium mass fraction at different ages. The black dashed line corresponds to the protosolar value at $Y=0.27$. At $t=2\,000~$Myr, the helium ocean is already there and starts growing. \textit{Bottom panel:} Temperature-pressure profiles of Saturn's envelope at these ages. Thick portions of the profiles indicate the helium ocean region. Phase curves from the SR2018 phase diagram with $T_{\rm offset}=350~$K are shown with grey lines for $Y=0.33$, 0.27, 0.18, 0.1 and 0.05. Grey-shaded areas correspond to the  de-mixing regions.}
         \label{figure:he-ocean}
\end{figure}

The potential formation of a helium ocean or helium core was already suggested by \citet{stevenson1977b} and recent models from \citet{pustow2016} and \citet{mankovich2020} have supported its presence in Saturn. Figure~\ref{figure:he-ocean} shows Saturn's evolution at different times (corresponding to the SR2018 case with $T_{\rm offset}=350~$K). The helium ocean appears $\sim$500~Myr after the onset of helium de-mixing. It then keeps growing throughout the evolution. The present-day structure of Saturn is therefore likely to harbour such a helium ocean. 
We argue that interior models designed to fit Saturn's gravity field and seismology data should account for the presence of a helium ocean. The thermal and electrical properties of a helium ocean have been investigated by \citep{preising2023}. We suggest that future studies investigating Saturn's magnetic field generation and interior dynamics should consider the existence of such a helium ocean.

%-------------------------------------------------------------------
%-------------------------------------------------------------------

%-------------------------------------------------------------------
%-------------------------------------------------------------------
\subsection{Model sensitivity to key ingredients}
  \label{subsection:effects}

\subsubsection{Equation of state}
\label{subsubsec:eos}

\begin{figure}[h]
   \centering
   \includegraphics[width=0.9\hsize]{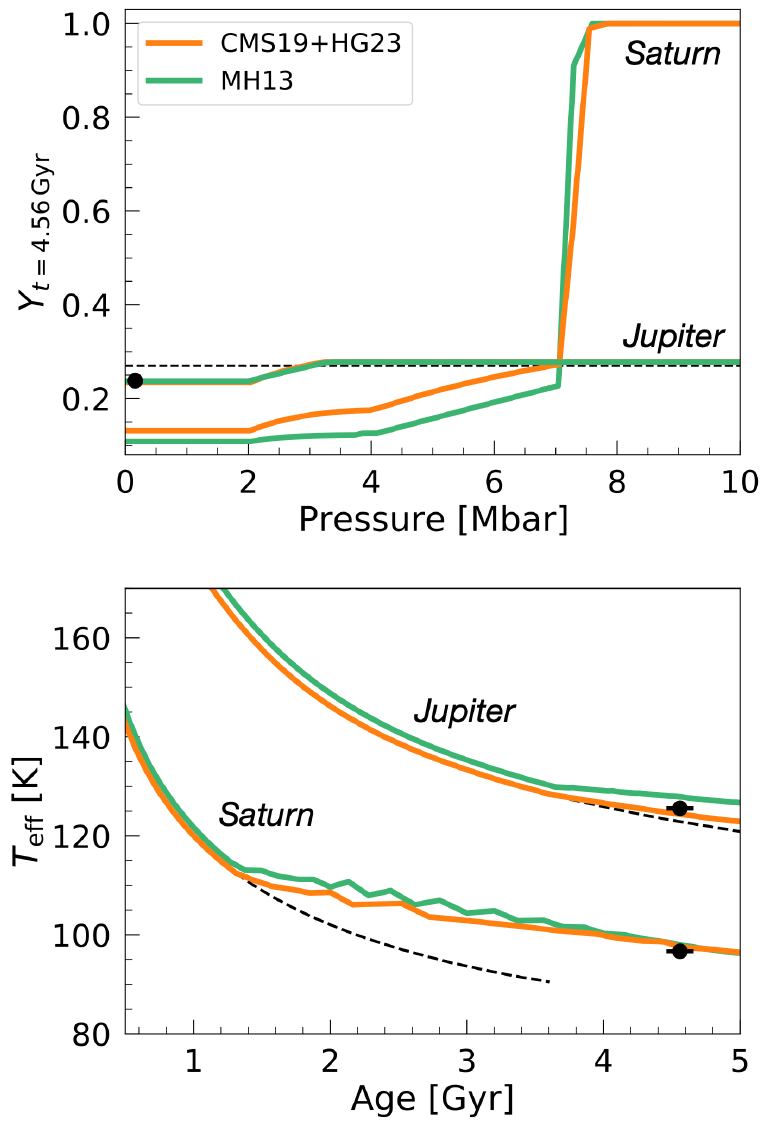}
      \caption{Effect of the EOS on the evolution of Jupiter and Saturn. We compare the CMS19+HG23 EOS \citep{chabrier2019,howard2023} and the MH13 EOS \citep{militzer2013}. The calculations use the SR2018 phase diagram with $T_{\rm offset}=350~$K (Sec~\ref{subsec:fiducial}). \textit{Top panel:} The present-day helium mass fraction as a function of pressure. The black dot shows the Galileo measured value. \textit{Bottom panel:}  Effective temperature as a function of age. The black  dashed lines show homogeneous evolution calculations. The black dots show the measured effective temperature and the errorbar corresponds to the age uncertainty.}
         \label{figure:eos_effect}
\end{figure}

In Sect.~\ref{subsec:fiducial}, we showed results using the CMS19+HG23 EOS. The non-ideal mixing effects due to interactions between hydrogen and helium are incorporated and depend on composition (see Sect.~\ref{subsec:model_setup}). With helium differentiation, it is crucial to properly assess the thermodynamical quantities as the composition changes with time. The helium mass fraction can span a wide range of values within the interiors of giant planets; in particular, in Saturn ($Y \in [0.1,1]$). Previous calculations used the MH13 EOS \citep{mankovich2020}. Such an EOS also includes non-ideal mixing effects, but for a constant composition ($Y=0.245$). Hence, we ran evolution calculations using the MH13 EOS and compared them to our results with CMS19+HG23. The comparison is shown in Fig.~\ref{figure:eos_effect}. We note that this is a specific case of the SR2018 phase diagram (with $T_{\rm offset}=350~$K). Results using the other phase diagrams are shown in Appendix~\ref{appendix:eos}. 

When we first tried to constrain the H-He phase diagrams by shifting them to fit Jupiter's evolution to the Galileo measurement, we found that the required temperature offsets are similar (different by less than 150~K). However, the effective temperatures obtained for Jupiter are significantly different from those obtained with CMS19+HG23. Before the onset of de-mixing, CMS19+HG23 yields a colder interior and therefore leads to a shorter cooling time of only about 0.2~Gyr compared to the case with the MH13 EOS.  Nonetheless, once de-mixing starts, the discrepancy between the two EOSs increases. Figure~\ref{figure:eos_effect} shows that while de-mixing starts at about the same age, the slope of the effective temperature with respect to age is different. The break of the curve from the homogeneous evolution is less pronounced with CMS19+HG23. Therefore, it seems that accounting for the composition-dependence of non-ideal mixing effects leads to a lower gain of energy from helium differentiation. The cooling time is thus shortened by about 1~Gyr. For Saturn, CMS19+HG23 yields larger $Y_1$ values (by 0.03 at most). As a result, models using MH13 are expected to underestimate the current helium content in Saturn's atmosphere. 

\subsubsection{Atmospheric model}
\label{subsubsec:atm_model}
\begin{figure}[h]
   \centering
   \includegraphics[width=0.9\hsize]{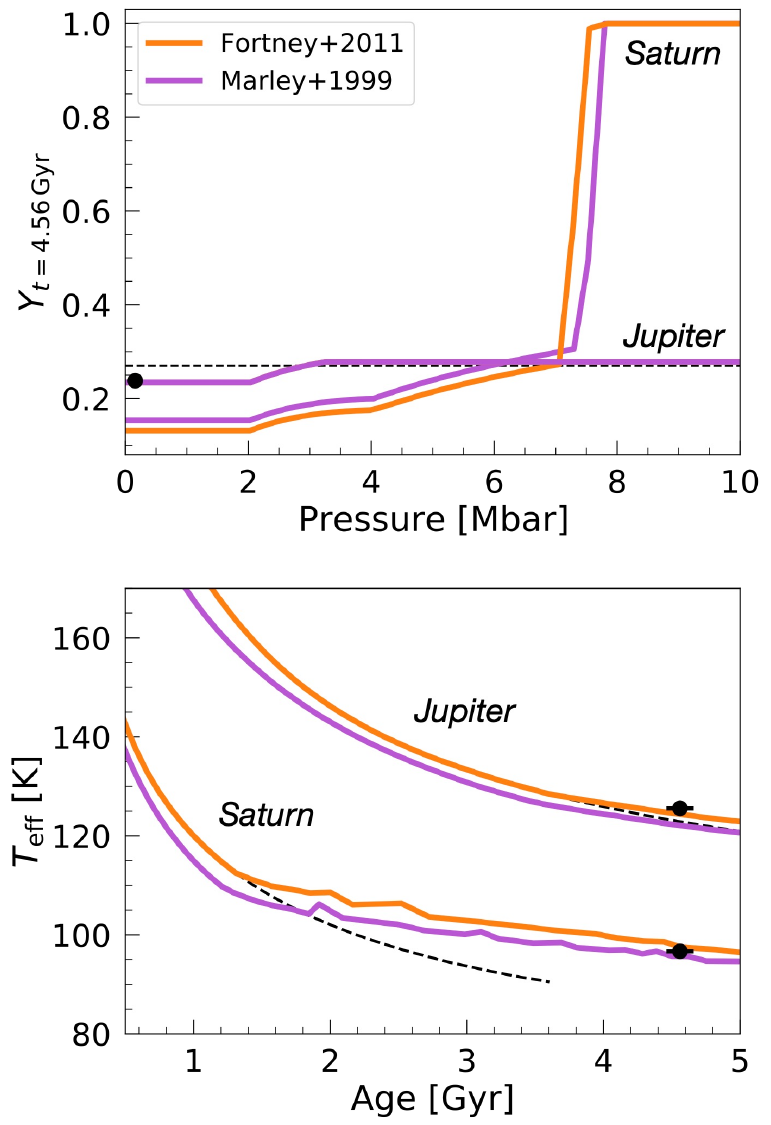}
      \caption{Effect of the atmospheric model on the evolution of Jupiter and Saturn. We compare the atmospheric models of \citet{fortney2011} and \citet{marley1999}. The calculations use the SR2018 phase diagram with $T_{\rm offset}=350~$K (Sec~\ref{subsec:fiducial}). \textit{Top panel:}  Present-day helium mass fraction as a function of pressure. The black dot shows the Galileo measured value. \textit{Bottom panel:}  Effective temperature as a function of age. The black dashed lines correspond to homogeneous evolution calculations while the black dots show the measured effective temperature and the errorbar gives the uncertainty in terms of age.}
         \label{figure:atm_effect}
\end{figure}

%Next we compare the results  {baseline?? canonical? } models using the atmospheric model from \citet{fortney2011} to calculations using the \citet{marley1999} model. 
Next, we investigated the sensitivity of the results to the used atmospheric models. We compare the results using the atmospheric model from \citet{fortney2011} to calculations using the \citet{marley1999} model. 
The latter is based on older opacity data and does not account for the evolution of the Sun's luminosity. The comparison is shown in Fig.~\ref{figure:atm_effect}. We found that the \citet{marley1999} atmospheric model leads to shorter cooling times, by at least 0.5~Gyr for Jupiter and up to almost 1~Gyr for Saturn. The constrained phase diagram applied to Saturn's evolution leads to a larger mass fraction of helium in the atmosphere, namely:\ by 0.02. The helium core is thus extended across a smaller range of pressures as less helium sinks to the deep interior. This comparison shows the significant effect of the atmospheric model on the evolution calculations of giant planets.

\subsubsection{Core mass}
\label{subsubsec:core_mass}

\begin{figure}[h]
   \centering
   \includegraphics[width=\hsize]{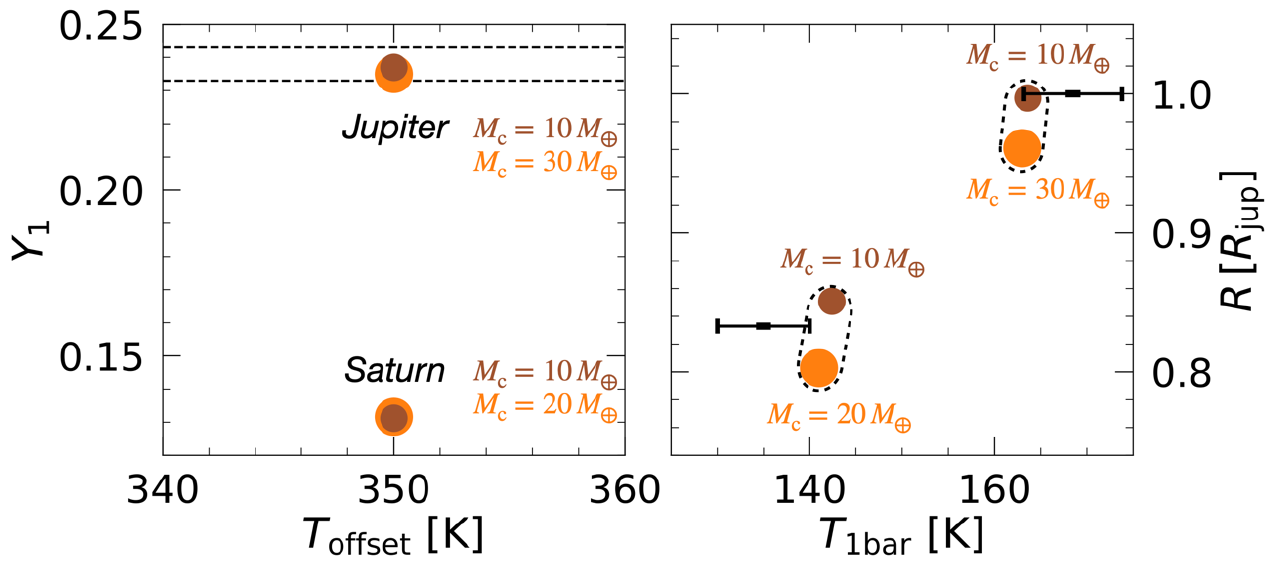}
      \caption{{Comparison of models with different core masses but when applying the same temperature offset. The calculations use the SR2018 phase diagram with $T_{\rm offset}=350~$K. Displayed quantities are at $t=4.56~$Gyr.
      \textit{Left panel:}  Atmospheric helium mass fraction as a function of the temperature offset. The horizontal dashed lines show the upper and lower bounds of the observed value of $Y_1$.
      \textit{Right panel:} Planetary radius as a function of the 1 bar temperature. The black errorbars show the observed radii and 1 bar temperatures of Jupiter and Saturn (see Fig.~\ref{figure:t1bar} for details about these measured values). The dashed lines show the possible range of models with core masses that fall between the two values we considered here.}}
      \label{figure:offsets_R_T1bar}
\end{figure}

\begin{figure}[h]
   \centering
   \includegraphics[width=0.9\hsize]{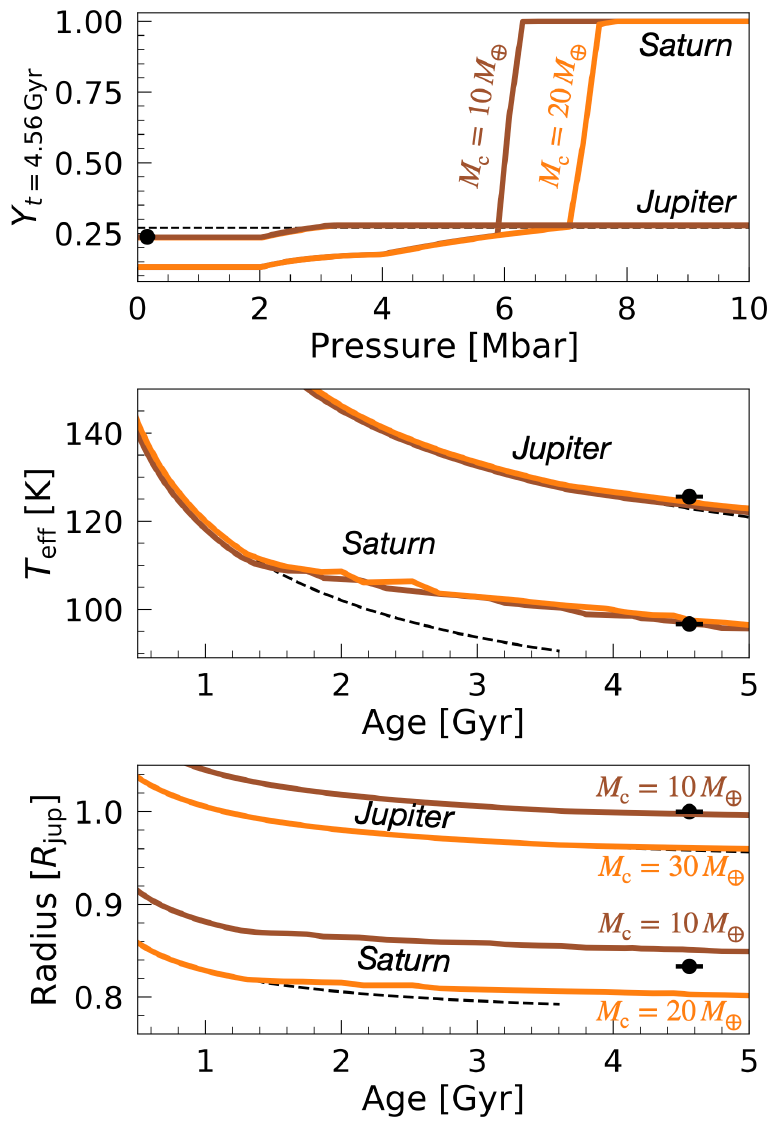}
      \caption{ Effect of the core mass on the evolution of Jupiter and Saturn. The calculations use the SR2018 phase diagram with $T_{\rm offset}=350~$K. \textit{Top panel:}  Present-day helium mass fraction as a function of pressure. The black dot shows the Galileo measured value. \textit{Middle panel:}  Effective temperature as a function of age. Black dots show the measured effective temperature. \textit{Bottom panel:}  Planet's radius as a function of age. The black dashed lines correspond to homogeneous evolution calculations. Black dots show the measured radii and the errorbar the uncertainty in age.}
         \label{figure:core_mass}
\end{figure}

As mentioned in Sect.~\ref{subsec:model_setup}, our baseline models used a core mass $M_{\rm c}$ of $30~M_{\oplus}$ for Jupiter and $20~M_{\oplus}$ for Saturn which is consistent with current interior models of the two planets but is still uncertain. Here we investigate the sensitivity of the results to the assumed core masses. We therefore also run calculations with a core mass of $10~M_{\oplus}$ for both planets. Using the SR2018 phase diagram {and applying $T_{\rm offset}=350~$K, Jupiter's model with a core mass of $10~M_{\oplus}$ yields $Y_1$=0.237 while the model with $30~M_{\oplus}$ yields $Y_1$=0.235 (Fig.~\ref{figure:offsets_R_T1bar}, left panel). Thus, a similar offset value allows both cases to match Jupiter's evolution to the observed atmospheric helium abundance. This suggests that our results for the temperature offsets are insensitive to the chosen core mass.
For Saturn, models with $M_{\rm c}=10~M_{\oplus}$ and $20~M_{\oplus}$ also yield similar values of $Y_1$. However, for both planets, models with different core masses yield different planetary radii (Fig.~\ref{figure:offsets_R_T1bar}, right panel). This shows that the observed planetary radius can be reconciled by adjusting the core mass, without considerably changing  the inferred temperature offset and atmospheric helium abundance. We also show the inferred 1 bar temperature of our models. Our Jupiter models agree with the lower bound of the measured $T_{\rm 1bar}$, while our Saturn models agree with the upper bound of the measurement. The values of $T_{\rm 1bar}$ obtained when applying the other phase diagrams are consistent with the measurements (see Appendix~\ref{appendix:t1bar}). In Appendix~\ref{appendix:t1bar}, we show how  $T_{\rm 1bar}$ (rather than time) can also be used to constrain the planetary evolution.}

For Jupiter, we find similar results for the helium distribution at $t=4.56~$Gyr for the two core mass values {(Fig.~\ref{figure:core_mass})}. However, for Saturn the inferred upper bound of the helium ocean is different. Decreasing $M_{\rm c}$ by 10~$M_{\oplus}$ leads to a present-day helium ocean from $P>6.3~$Mbar instead of 7.5~Mbar. This is because the presence of a more massive heavy-element core leads to higher pressure values in the planetary center. 
%\sh{The upper bound of the helium ocean is hence located at about 20\% of the planet's mass instead of 30\%.}
As a result, we can conclude that the heavy-element distribution in Saturn affects the location of a helium ocean. In addition, the effective temperatures yielded by calculations with a smaller core are slightly lower {(by less than 1~K, in agreement with \citet{mankovich2016})}, as these models start colder because they include a smaller mass of heavy elements. The evolution with a core mass of 10~$M_{\oplus}$ agrees with Jupiter's mean radius while a core mass of between 10~$M_{\oplus}$ and 20~$M_{\oplus}$ would agree with Saturn's mean radius. Fitting the planets' radii would hence require a core mass lower than realistic total heavy-element mass inferred from interior models based on the measured gravitational field (and seismology in the case of Saturn) data \citep{mankovich2021,miguel2022,howard2023_interior}, especially for Jupiter. {This test demonstrates that adjusting the core mass (i.e., the heavy-element content) could yield the measured radius without affecting significantly the required $T_{\rm offset}$ and inferred $Y_1$ and $T_{\rm eff}$ values at present-day.}

\subsubsection{Super-adiabaticity}
\label{subsubsec:superadiabaticity}

As helium rains out, a helium gradient forms in the interiors of Jupiter and Saturn. This helium gradient could inhibit convective mixing.  Double-diffusion \citep{rosenblum2011,mirouh2012,wood2013} may occur, resulting in a super-adiabatic temperature gradient. We parameterize the level of super-adiabaticity in the region of the helium gradient using the density ratio, $R_{\rho}$. The temperature gradient can then be defined as \citep[see also][]{mankovich2020,nettelmann2015}: 
\begin{equation}
    \nabla_{\rm T}=\nabla_{\rm ad}+R_{\rho} B,
    \label{eq:ledoux}
\end{equation}
with
\begin{equation}
    B = \frac{\chi_{\rho}}{\chi_{\rm T}}
    \left(\frac{d\,\textrm{ln}\,\rho}{d\, \textrm{ln}\,Y} \right)_{P,T} \nabla_{Y}, 
    \label{eq:composition_term}
\end{equation}
where 
\begin{equation}
    \chi_{\rho} = \left(\frac{d\,\textrm{ln}\,P}{d\, \textrm{ln}\,\rho} \right)_{T,Y} ; \qquad 
    \chi_{\rm T} = \left(\frac{d\,\textrm{ln}\,P}{d\, \textrm{ln}\,T} \right)_{\rho,Y} ; \qquad 
    \nabla_{Y}=\frac{d\,\textrm{ln}\,Y}{d\,\textrm{ln}\,P}.
    \label{eq:chi_terms}
\end{equation}
Therefore, the expected change in temperature due to super-adiabaticity depends on the helium gradient and also on the EOS as $\chi_{\rho}$ and $\chi_{\rm T}$ are inferred from the EOS tables. 
We use $R_{\rho}=0.05$ and $R_{\rho}=0.1$ for Jupiter and $R_{\rho}=0.004$ and $R_{\rho}=0.008$ for Saturn and simulate their evolution with these values. This time, we used the Lorenzen2011 phase diagram (with $T_{\rm offset}=-1250~$K), for two reasons. First, it allowed us to find solutions for Saturn. Second, it ensured numerical stability as calculating the helium distribution becomes more challenging when considering super-adiabaticity. Furthermore, super-adiabatic models for Saturn have been smoothed with splines to remove spurious data and focus on the main trend of the evolution. 
%{FIX: We acknowledge that these results are inherently influenced by the fitting process, thus warranting a cautious interpretation.  } 

\begin{figure}[h]
   \centering
   \includegraphics[width=0.9\hsize]{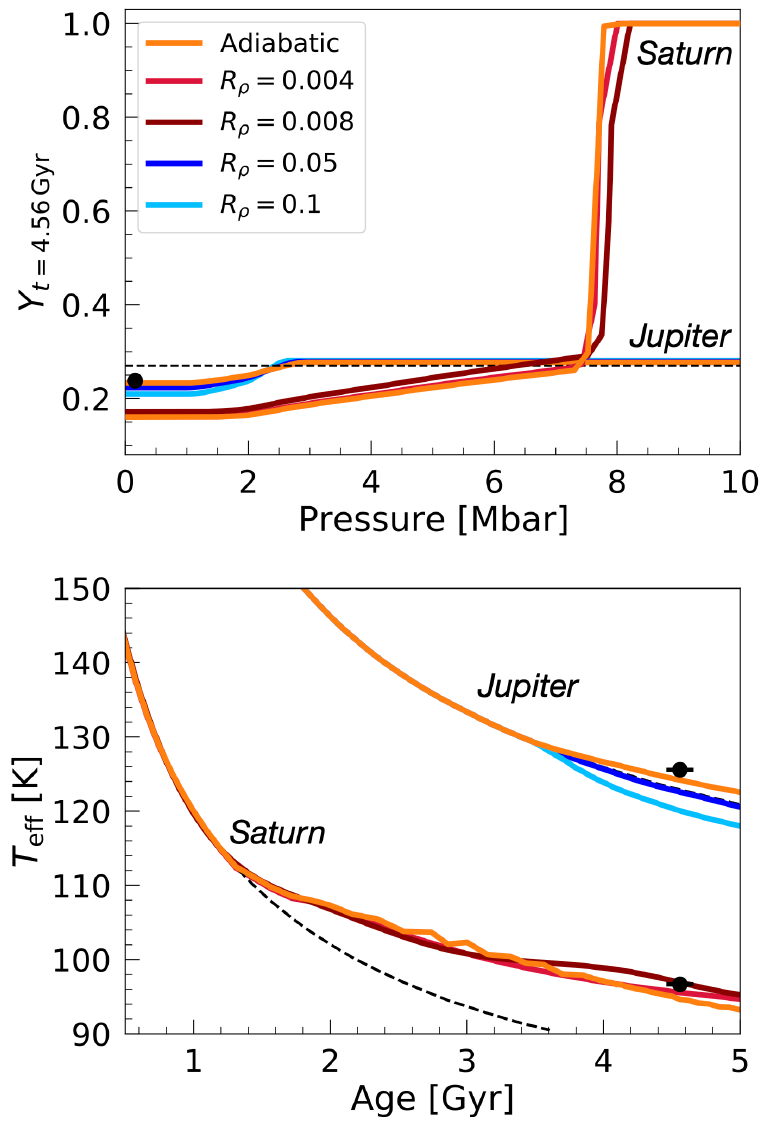}
      \caption{Effect of super-adiabaticity on the thermal evolution of Jupiter and Saturn. We compare calculations with different values of $R_{\rho}$ (see Eq.~\ref{eq:ledoux}). The simulations use the Lorenzen2011 phase diagram with $T_{\rm offset}=-1250~$K. \textit{Top panel:}  Present-day helium mass fraction as a function of pressure. The black dot shows the Galileo measured value. \textit{Bottom panel:} Effective temperature as a function of age. The black dashed lines correspond to homogeneous evolution calculations, while the black dots show the measured effective temperature and the errorbar the uncertainty in age.}
         \label{figure:superad1}
\end{figure}

The simulation results are shown in Fig.~\ref{figure:superad1}. As expected, we find that a super-adiabatic region accelerates the cooling of Jupiter and hence yields lower $Y_1$ values \citep[see, e.g.,][]{mankovich2016,sur2024}. 
Our baseline model (adiabatic) already yielded an effective temperature slightly lower than the measured one, as a result, the inclusion of a super-adiabatic temperature gradient due to the helium gradient does not help reconcile the measurement. However, given the sources of uncertainties discussed previously (EOS, atmospheric model), we cannot exclude the existence of such a super-adiabatic region in Jupiter. {Previous works suggested so \citep{fortney2003,nettelmann2015,mankovich2016}. In particular,} \citet{mankovich2020} derived a  modest super-adiabaticity ($R_{\rho} \sim 0.05$, although the number comparison is not straightforward because  $R_{\rho}$ depends on the EOS and the $Y$ gradient) using MH13 instead of CMS19+HG23, which may have overestimated Jupiter's cooling time. {When using MH13 (see Appendix~\ref{appendix:eos}), we found $T_{\rm eff}$ larger than the measured value, allowing us to also infer $R_{\rho}$ that is on the same order of magnitude as in \citet{mankovich2020}. However, the improved CMS19+HG23 EOS prevents us from finding models with a super-adiabatic gradient, as this would lead to lower $T_{\rm eff}$.} The presence of heavy elements in the envelope in the models of \citet{mankovich2020} also leads to a hotter interior.  
For Saturn (as discussed above) helium rain begins earlier. We find that super-adiabatic models first accelerate the cooling once de-mixing begins. However,  after a certain time ($\sim 2~$Gyr), the planetary cooling and contraction are delayed. This is because heat is initially stored in the planetary deep interior, which leads in the early stages after the onset of de-mixing to a lower $T_{\rm eff}$ compared to the adiabatic case \citep[e.g.,][]{leconte2013}. Heat is then released in the outer part of the planet, which leads to a larger $T_{\rm eff}$ at present.  As a result, these Saturn evolution models coupled with Lorenzen2011 and that include super-adiabaticity are more consistent with the measured effective temperature. On the other hand, the SR2018 phase diagram led to slightly higher $T_{\rm eff}$ with adiabatic models (see Fig.~\ref{figure:Saturn_HG23}). In this case, super-adiabatic models would appear less consistent with the effective temperature measurement. We note that \citet{mankovich2020} found that super-adiabatic ($R_{\rho}<0.065$) and adiabatic models are equally likely for Saturn. We confirm that it is hard to draw clear conclusions regarding the magnitude of super-adiabaticity in Saturn's interior.

\begin{figure}[h]
   \centering
   \includegraphics[width=0.9\hsize]{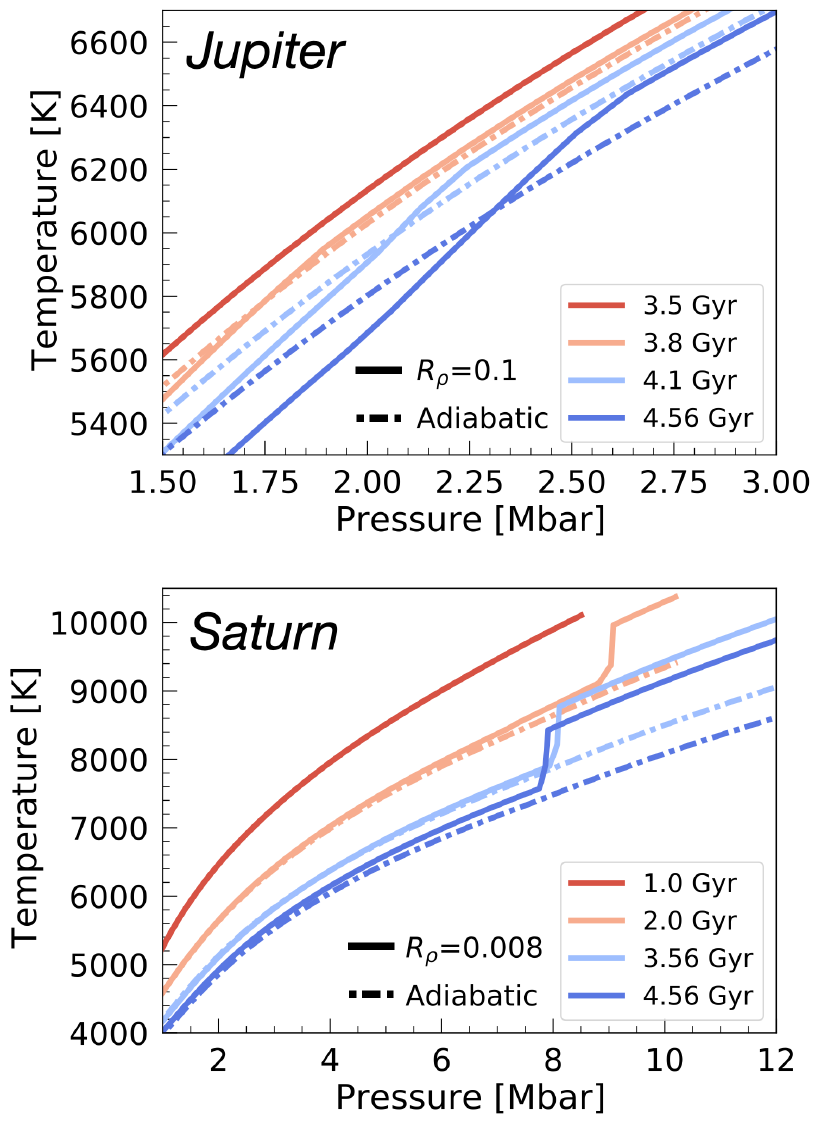}
      \caption{Temperature-pressure profiles of the envelopes of Jupiter and Saturn at different ages, for adiabatic and super-adiabatic interiors. We compare calculations shown on Fig.~\ref{figure:superad1}. Dash-dotted lines show $T-P$ profiles of our baseline models, assuming adiabaticity ($R_{\rho}=0$). Solid lines show $T-P$ profiles for $R_{\rho}=0.1$ for Jupiter (\textit{top panel}) and $R_{\rho}=0.008$ for Saturn (\textit{bottom panel}).}
         \label{figure:superad2}
\end{figure}

Figure ~\ref{figure:superad2} shows the temperature-pressure profiles of the adiabatic and super-adiabatic evolution models of Jupiter and Saturn. One can hence relate the density ratio $R_{\rho}$ to the increase of temperature in the planetary deep interiors. At $t=4.56~$Gyr, the temperature in Jupiter's outer region is lower when $R_{\rho}=0.1$ compared to the adiabatic case. On the other hand, in Saturn's outer region, the temperature at the same age is higher when $R_{\rho}=0.008$ compared to the adiabatic case. This confirms the opposite behaviour of Jupiter and Saturn: super-adiabatic models result in a lower $T_{\rm eff}$ for Jupiter today, while they yield a higher $T_{\rm eff}$ for Saturn today. Nevertheless, constraining the exact level of super-adiabaticity in Jupiter and Saturn remains challenging with such evolutionary models. The presence of heavy elements in the envelope and the existence of a fuzzy core would also affect the long-term thermal evolution of the planets. Measuring the helium content in Saturn's atmosphere by a probe could help constrain the evolution of Saturn and put limits on the magnitude of the super-adiabaticity of the helium gradient.

\section{Discussion}
\label{section:4}

\subsection{Limitations}
While our study presents a step forward towards a better understanding of the evolution of Jupiter and Saturn with helium rain, it also has some limitations. For simplicity, we assumed that all the heavy elements are concentrated in the planets' cores and the envelopes consist of a pure H-He mixture. Such a heavy-element distribution is oversimplified. First, homogeneous envelopes (even with $Z \neq 0$) are at odds with updated interior models of the planets that infer the presence of heavy-element gradients in the deep interiors of both Jupiter and Saturn \citep[e.g.,][]{wahl2017,debras2019,mankovich2021,militzer2022,howard2023_interior}. 
Second, the atmospheres of Jupiter and Saturn are expected to be enriched in heavy elements by a factor of $\sim3$ and $\sim 5-10$ compared to protosolar for Jupiter and Saturn, respectively \citep[see e.g.,][and references therein]{guillot2023}.
However, the simplified assumed internal structure allows us to consistently couple our evolution models in which the planetary envelopes consist of pure H-He, consistent with available phase diagrams that were constructed for pure H-He mixtures. 
Since our models include simplifications,  the inferred required temperature offsets, $T_{\rm offset}$, and the predicted mass fractions of helium in the outer ($Y_1$) and inner ($Y_2$) parts of the planetary interior should be taken as indicative of a trend. 
\par

We note that constraining the exact level of super-adiabaticity in the formed helium gradient in Jupiter and Saturn remains hard with evolution calculations. Our models suggest that Saturn's helium gradient could be super-adiabatic. However, the potential presence of a heavy-element gradient could significantly affect the thermal evolution of both planets \citep{leconte2013,vazan2016}. This should be studied further and implemented in future evolution models of Jupiter and Saturn. Theoretical work on convective inhibition \citep{guillot1995_condensation} could also help to constrain the properties of the super-adiabatic temperature gradient in the phase separating region. 
%{Recently}, studies of the inhibition of convection focused on condensation in hydrogen atmospheres \citep{leconte2017,markham2021} and oceans \citep{markham2022}. 
{\citet{nettelmann2015} investigated the influence of convection inhibition due to H-He phase separation on Jupiter's temperature profile and heat transport and constrained the height of semiconvective layers. Future studies could consider magnetic field measurements, which seem to support the presence of a stably stratified layer near the region where phase separation occurs \citep{connerney2022,moore2022}.}

Finally, at the moment, the helium atmospheric abundance in Saturn is uncertain. In this work, we found that using the MH13 EOS \citep{militzer2013} instead of CMS19+HG23 \citep{chabrier2019,howard2023} leads to $Y_1$ values reduced to 0.10. A probe to Saturn \citep{mousis2016,fortney2023} that will measure the helium abundance in the atmosphere could test our model predictions, and constrain Saturn's evolution and its internal structure. 
%\qsh{why not in the conclusion? (6?)}

%In addition, other complications for instance by accounting rotation which affects the shape of the planets have not been considered. 
%Modern and denser H-He EOSs make it challenging for evolution calculations assuming well separated core and envelope to yield the correct radius. In these simulations we did not include heavy elements in the envelope, which would make the situation worse. 
%We hence did not intend to fit the planet's radii.

\subsection{Microphysics and atmospheres}
The evolution models we presented strongly rely on the EOS which is uncertain. Using a state-of-the-art EOS for H-He yields colder planets. Accounting for non-ideal mixing effects, which depend on the helium fraction (CMS19+HG23 EOS, \citet{howard2023}) leads to an energy gain from de-mixing lower than what is obtained with mixing effects for constant composition (MH13 EOS, \citet{militzer2013}). The cooling time can be reduced by several hundred million years in the case of Jupiter when de-mixing has started. The atmospheric model also has an important effect. Using updated atmospheric models, accounting for the evolution of the Sun's luminosity and Jupiter's revised Bond albedo, we find longer cooling times than previous atmospheric models, on the order of a few hundred million years. Further investigations of the physical and chemical processes occurring in giant planet atmospheres and the implementation of more realistic atmospheric models \citep[e.g.,][]{chen2023} will allow for a better understanding of the thermal evolution of giant planets. \par

An improved understanding of the H-He phase diagram is clearly desirable to further constrain the planetary evolution. At the moment, there is still a large discrepancy between various theoretical calculations and the single experiment result. While our results for the evolution of Jupiter and Saturn provide hints on the location and shape of the exact phase diagram, further explorations of the immiscibility between hydrogen and helium are needed, across a wide range of pressures and temperatures, and for several mixture ratios. In addition, future ab initio simulations and experiments that incorporate heavy elements are required. For example, the presence of helium, even in small quantities, delays the metallization of hydrogen \citep{vorberger2007,mazzola2018,helled2020}. The inclusion of heavy elements in theoretical calculations and experiments could therefore influence the phase separation between hydrogen and helium. This should be investigated and implemented in evolution models. 

\section{Summary and conclusions}
\label{section:5}

We presented thermal evolution models of Jupiter and Saturn accounting for helium rain. We applied theoretical \citep{lorenzen2011,schottler2018_prl} H-He phase diagrams {and our construction of a phase diagram based on the experimental data of \citet{brygoo2021}} and investigated their effect on the evolution of both planets. Our models of Jupiter and Saturn assumed a central dense core surrounded by a H-He envelope consistent with current H-He phase diagrams.
Overall, the main conclusions from this study {depend on our assumptions (e.g., the applied H-He EOS) and} can be summarized as follows: 

\begin{enumerate}
    \item An offset in temperature of -1250, +350, and -3850~K, respectively, is required for the phase diagrams of \citet{lorenzen2011}, \citet{schottler2018_prl} and \citet{brygoo2021} to find agreement between Jupiter's evolution and the Galileo measurement of its atmospheric helium abundance ($Y_{\rm atm}=0.238 \pm 0.005$). 
    \item When applying the same offset in temperature to the evolution of both Jupiter and Saturn, our models of both planets are in rather good agreement with their measured effective temperature {(by less than 2~K)}. 
    This suggests that the exact H-He phase diagram may be obtained by shifting the existing ones with our estimated offsets in temperature (see Fig.~\ref{figure:phase_diagrams}).
    \item The temperature shift implied by the experimental phase diagram \citep{brygoo2021} is substantial (-3850~K), suggesting that the original experimental data is very inconsistent with the evolution of both planets. It would lead to {a substantial} depletion of helium ($Y_1 \sim 0.04$) in Jupiter's atmosphere in less than 2~Gyr.
    \item Demixing starts at about 3.5~Gyr for Jupiter and at about 1.3~Gyr for Saturn. Helium rain leads to very different interiors between Jupiter and Saturn. Only slight de-mixing ($Y_1=0.238$) in a small part of Jupiter (between 1 and 3~Mbar) occurs, whereas it leads to the presence of a large helium gradient (from $Y_1 \sim 0.13$ to $Y_2 \sim 1$)  on a large proportion of Saturn (between 2 and 9~Mbar). 
    \item  Saturn is likely to have a helium ocean which forms rapidly in a few hundred million years after the onset of He de-mixing. {We find $Y \sim 1$ in this region, due to simplifying assumptions; however, more realistic values are likely to be around $0.9-0.95$ \citep{pustow2016,mankovich2020}.} The extent of this helium ocean depends on the size of the pure heavy-element core. 
    \item Saturn's atmospheric helium mass fraction is between 0.13 and 0.16, consistent with recent Cassini's estimates, namely: the lower bound of \citet{koskinen2018} (0.16--0.22) and the upper bound of \citet{achterberg2020} (0.075--0.13). {However, measurements of Saturn's atmospheric helium content cover a broad range of values (see Fig.~\ref{figure:t1bar}) and a more accurate determination is still required.} 
\end{enumerate}

{Our evolution models confirm and are in agreement with previous results. 
%The values of the required temperature offsets found are more or less similar when using the same phase diagram and considering that we did not use the same H-He EOSs. 
In this work, we find consistent values for the atmospheric helium abundance and similar ages for the onset of helium rain. These findings demonstrate the robustness of our evolution calculations.
} Overall, evolution models serve as significant complements to static interior models and can unveil important information on the planetary structure, for example, the presence of a helium ocean in Saturn. 
Future evolution models, including both heavy-element gradients \citep{vazan2018,muller2020} and H-He phase separation will provide new insights into the evolution and internal structure of giant planets. 
\par 
Our comprehension of giant planet interiors will also benefit from an improved 
%from advancements in various paths of research. Progress in  
understanding of the behaviour of hydrogen and helium at high pressures and temperatures, which yields accurate phase diagrams and EOSs. 
In addition, measuring Saturn's atmospheric composition by an entry probe  {is needed to} determine the He abundance in Saturn’s atmosphere, constrain the magnitude of He rain, and provide additional clues on the H-He phase diagram and therefore on its formation and evolution history.
%This will also allow a direct comparsion with Jupiter. 
%and determining Saturn’s atmospheric metallicity. 
%is crucial for interior and evolution models. 
%Interior models of Jupiter and Saturn, based on gravity and seismology data, are obviously key. Ongoing efforts to reconcile these models with measurements are currently underway \citep{howard2023_invZ,muller2024}. 
Furthermore, the detection of warm and cold giant exoplanets allows us to apply our knowledge from the Solar system's giant planets to characterize giant planets around other stars \citep{muller2023}. Finally, further synergies between planetary science, high-pressure physics, and exoplanetary science, as well as advancements from observations, laboratory experiments, and theoretical and numerical calculations will {be essential to} reveal the nature of gas giant planets in our planetary system and beyond.

%{In addition, our understanding of the connection between the atmosphere and deep interior of giant planets is still incomplete. Bulk composition and atmpospheric composition... There are stills tensions to explain the observed atmospheric enrichment \citep{howard2023_invZ,muller2024}. Future evolution models, including heavy-element gradients may provide new insights, although it comes with additional uncertainties: how this heavy-element gradient forms and evolves \citep{vazan2018,muller2020} and how heat is transported there.}

%{Happy ending :)}

%-------------------------------------------------------------------
%-------------------------------------------------------------------

\begin{acknowledgements}
{We thank the referee for valuable comments which helped improve the manuscript.}
We acknowledge support from SNSF grant \texttt{\detokenize{     200020_215634}} and the National Centre for Competence in Research ‘PlanetS’ supported by SNSF. We thank the Juno team for insightful discussions. We also thank Christopher Mankovich for helpful discussions.
\end{acknowledgements}

% WARNING
%-------------------------------------------------------------------
% Please note that we have included the references to the file aa.dem in
% order to compile it, but we ask you to:
%
% - use BibTeX with the regular commands:
   \bibliographystyle{aa} % style aa.bst
   \bibliography{biblio} % your references Yourfile.bib

\begin{thebibliography}{80}
\expandafter\ifx\csname natexlab\endcsname\relax\def\natexlab#1{#1}\fi

\bibitem[{{Achterberg} \& {Flasar}(2020)}]{achterberg2020}
{Achterberg}, R.~K. \& {Flasar}, F.~M. 2020, The Planetary Science Journal, 1, 30

\bibitem[{{Asplund} {et~al.}(2021){Asplund}, {Amarsi}, \& {Grevesse}}]{asplund2021}
{Asplund}, M., {Amarsi}, A.~M., \& {Grevesse}, N. 2021, \aap, 653, A141

\bibitem[{{Brygoo} {et~al.}(2021){Brygoo}, {Loubeyre}, {Millot}, {Rygg}, {Celliers}, {Eggert}, {Jeanloz}, \& {Collins}}]{brygoo2021}
{Brygoo}, S., {Loubeyre}, P., {Millot}, M., {et~al.} 2021, \nat, 593, 517

\bibitem[{{Chabrier} {et~al.}(2019){Chabrier}, {Mazevet}, \& {Soubiran}}]{chabrier2019}
{Chabrier}, G., {Mazevet}, S., \& {Soubiran}, F. 2019, \apj, 872, 51

\bibitem[{{Chen} {et~al.}(2023){Chen}, {Burrows}, {Sur}, \& {Arevalo}}]{chen2023}
{Chen}, Y.-X., {Burrows}, A., {Sur}, A., \& {Arevalo}, R.~T. 2023, \apj, 957, 36

\bibitem[{Connelly {et~al.}(2012)Connelly, Bizzarro, Krot, Åke Nordlund, Wielandt, \& Ivanova}]{connely2012}
Connelly, J.~N., Bizzarro, M., Krot, A.~N., {et~al.} 2012, Science, 338, 651

\bibitem[{{Connerney} {et~al.}(2022){Connerney}, {Timmins}, {Oliversen}, {Espley}, {Joergensen}, {Kotsiaros}, {Joergensen}, {Merayo}, {Herceg}, {Bloxham}, {Moore}, {Mura}, {Moirano}, {Bolton}, \& {Levin}}]{connerney2022}
{Connerney}, J.~E.~P., {Timmins}, S., {Oliversen}, R.~J., {et~al.} 2022, Journal of Geophysical Research (Planets), 127, e07055

\bibitem[{{Conrath} \& {Gautier}(2000)}]{conrath2000}
{Conrath}, B.~J. \& {Gautier}, D. 2000, \icarus, 144, 124

\bibitem[{{Conrath} {et~al.}(1984){Conrath}, {Gautier}, {Hanel}, \& {Hornstein}}]{conrath1984}
{Conrath}, B.~J., {Gautier}, D., {Hanel}, R.~A., \& {Hornstein}, J.~S. 1984, \apj, 282, 807

\bibitem[{{Debras} \& {Chabrier}(2019)}]{debras2019}
{Debras}, F. \& {Chabrier}, G. 2019, \apj, 872, 100

\bibitem[{{Fortney} \& {Hubbard}(2003)}]{fortney2003}
{Fortney}, J.~J. \& {Hubbard}, W.~B. 2003, \icarus, 164, 228

\bibitem[{{Fortney} {et~al.}(2011){Fortney}, {Ikoma}, {Nettelmann}, {Guillot}, \& {Marley}}]{fortney2011}
{Fortney}, J.~J., {Ikoma}, M., {Nettelmann}, N., {Guillot}, T., \& {Marley}, M.~S. 2011, \apj, 729, 32

\bibitem[{{Fortney} {et~al.}(2023){Fortney}, {Militzer}, {Mankovich}, {Helled}, {Wahl}, {Nettelmann}, {Hubbard}, {Stevenson}, {Iess}, {Marley}, \& {Movshovitz}}]{fortney2023}
{Fortney}, J.~J., {Militzer}, B., {Mankovich}, C.~R., {et~al.} 2023, arXiv e-prints, arXiv:2304.09215

\bibitem[{{Fortney} \& {Nettelmann}(2010)}]{fortneynettelmann2010}
{Fortney}, J.~J. \& {Nettelmann}, N. 2010, \ssr, 152, 423

\bibitem[{{Graboske} {et~al.}(1975){Graboske}, {Pollack}, {Grossman}, \& {Olness}}]{graboske1975}
{Graboske}, H.~C., J., {Pollack}, J.~B., {Grossman}, A.~S., \& {Olness}, R.~J. 1975, \apj, 199, 265

\bibitem[{{Guillot}(1995)}]{guillot1995_condensation}
{Guillot}, T. 1995, Science, 269, 1697

\bibitem[{{Guillot}(2005)}]{guillot2005}
{Guillot}, T. 2005, Annual Review of Earth and Planetary Sciences, 33, 493

\bibitem[{{Guillot} {et~al.}(1995){Guillot}, {Chabrier}, {Gautier}, \& {Morel}}]{guillot1995_EvRad}
{Guillot}, T., {Chabrier}, G., {Gautier}, D., \& {Morel}, P. 1995, \apj, 450, 463

\bibitem[{{Guillot} {et~al.}(2023){Guillot}, {Fletcher}, {Helled}, {Ikoma}, {Line}, \& {Paramentier}}]{guillot2023}
{Guillot}, T., {Fletcher}, L.~N., {Helled}, R., {et~al.} 2023, in Astronomical Society of the Pacific Conference Series, Vol. 534, Protostars and Planets VII, ed. S.~{Inutsuka}, Y.~{Aikawa}, T.~{Muto}, K.~{Tomida}, \& M.~{Tamura}, 947

\bibitem[{{Guillot} \& {Morel}(1995)}]{guillot1995_cepam}
{Guillot}, T. \& {Morel}, P. 1995, \aaps, 109, 109

\bibitem[{{Gupta} {et~al.}(2022){Gupta}, {Atreya}, {Steffes}, {Fletcher}, {Guillot}, {Allison}, {Bolton}, {Helled}, {Levin}, {Li}, {Lunine}, {Miguel}, {Orton}, {Hunter Waite}, \& {Withers}}]{gupta2022}
{Gupta}, P., {Atreya}, S.~K., {Steffes}, P.~G., {et~al.} 2022, \psj, 3, 159

\bibitem[{{Helled}(2018)}]{helled2018}
{Helled}, R. 2018, in Oxford Research Encyclopedia of Planetary Science, 175

\bibitem[{{Helled} \& {Howard}(2024)}]{RavitSaburo}
{Helled}, R. \& {Howard}, S. 2024, arXiv e-prints, arXiv:2407.05853

\bibitem[{{Helled} {et~al.}(2020){Helled}, {Mazzola}, \& {Redmer}}]{helled2020}
{Helled}, R., {Mazzola}, G., \& {Redmer}, R. 2020, Nature Reviews Physics, 2, 562

\bibitem[{{Howard} \& {Guillot}(2023)}]{howard2023}
{Howard}, S. \& {Guillot}, T. 2023, \aap, 672, L1

\bibitem[{{Howard} {et~al.}(2023){Howard}, {Guillot}, {Bazot}, {Miguel}, {Stevenson}, {Galanti}, {Kaspi}, {Hubbard}, {Militzer}, {Helled}, {Nettelmann}, {Idini}, \& {Bolton}}]{howard2023_interior}
{Howard}, S., {Guillot}, T., {Bazot}, M., {et~al.} 2023, \aap, 672, A33

\bibitem[{{Hubbard}(1969)}]{hubbard1969}
{Hubbard}, W.~B. 1969, \apj, 155, 333

\bibitem[{{Hubbard}(1977)}]{hubbard1977}
{Hubbard}, W.~B. 1977, \icarus, 30, 305

\bibitem[{{Hubbard} \& {Dewitt}(1985)}]{HDW1985}
{Hubbard}, W.~B. \& {Dewitt}, H.~E. 1985, \apj, 290, 388

\bibitem[{{Hubbard} {et~al.}(1999){Hubbard}, {Guillot}, {Marley}, {Burrows}, {Lunine}, \& {Saumon}}]{hubbard1999}
{Hubbard}, W.~B., {Guillot}, T., {Marley}, M.~S., {et~al.} 1999, \planss, 47, 1175

\bibitem[{{Koskinen} \& {Guerlet}(2018)}]{koskinen2018}
{Koskinen}, T.~T. \& {Guerlet}, S. 2018, \icarus, 307, 161

\bibitem[{{Leconte} \& {Chabrier}(2013)}]{leconte2013}
{Leconte}, J. \& {Chabrier}, G. 2013, Nature Geoscience, 6, 347

\bibitem[{{Ledoux}(1947)}]{ledoux1947}
{Ledoux}, P. 1947, \apj, 105, 305

\bibitem[{Li {et~al.}(2018)Li, Jiang, West, Gierasch, Perez-Hoyos, Sanchez-Lavega, Fletcher, Fortney, Knowles, Porco, Baines, Fry, Mallama, Achterberg, Simon, Nixon, Orton, Dyudina, Ewald, \& Schmude}]{li2018}
Li, L., Jiang, X., West, R.~A., {et~al.} 2018, Nature Communications, 9, 3709

\bibitem[{{Lindal} {et~al.}(1985){Lindal}, {Sweetnam}, \& {Eshleman}}]{lindal1985}
{Lindal}, G.~F., {Sweetnam}, D.~N., \& {Eshleman}, V.~R. 1985, \aj, 90, 1136

\bibitem[{Lorenzen {et~al.}(2009)Lorenzen, Holst, \& Redmer}]{lorenzen2009}
Lorenzen, W., Holst, B., \& Redmer, R. 2009, Phys. Rev. Lett., 102, 115701

\bibitem[{Lorenzen {et~al.}(2011)Lorenzen, Holst, \& Redmer}]{lorenzen2011}
Lorenzen, W., Holst, B., \& Redmer, R. 2011, Phys. Rev. B, 84, 235109

\bibitem[{{Low}(1966)}]{low1966}
{Low}, F.~J. 1966, \aj, 71, 391

\bibitem[{{Mankovich} {et~al.}(2016){Mankovich}, {Fortney}, \& {Moore}}]{mankovich2016}
{Mankovich}, C., {Fortney}, J.~J., \& {Moore}, K.~L. 2016, \apj, 832, 113

\bibitem[{{Mankovich} \& {Fortney}(2020)}]{mankovich2020}
{Mankovich}, C.~R. \& {Fortney}, J.~J. 2020, \apj, 889, 51

\bibitem[{{Mankovich} \& {Fuller}(2021)}]{mankovich2021}
{Mankovich}, C.~R. \& {Fuller}, J. 2021, Nature Astronomy, 5, 1103

\bibitem[{{Marley} {et~al.}(1999){Marley}, {Gelino}, {Stephens}, {Lunine}, \& {Freedman}}]{marley1999}
{Marley}, M.~S., {Gelino}, C., {Stephens}, D., {Lunine}, J.~I., \& {Freedman}, R. 1999, \apj, 513, 879

\bibitem[{{Mazzola} {et~al.}(2018){Mazzola}, {Helled}, \& {Sorella}}]{mazzola2018}
{Mazzola}, G., {Helled}, R., \& {Sorella}, S. 2018, \prl, 120, 025701

\bibitem[{{Miguel} {et~al.}(2022){Miguel}, {Bazot}, {Guillot}, {Howard}, {Galanti}, {Kaspi}, {Hubbard}, {Militzer}, {Helled}, {Atreya}, {Connerney}, {Durante}, {Kulowski}, {Lunine}, {Stevenson}, \& {Bolton}}]{miguel2022}
{Miguel}, Y., {Bazot}, M., {Guillot}, T., {et~al.} 2022, \aap, 662, A18

\bibitem[{{Miguel} \& {Vazan}(2023)}]{miguel2023}
{Miguel}, Y. \& {Vazan}, A. 2023, Remote Sensing, 15, 681

\bibitem[{{Militzer} \& {Hubbard}(2013)}]{militzer2013}
{Militzer}, B. \& {Hubbard}, W.~B. 2013, \apj, 774, 148

\bibitem[{{Militzer} {et~al.}(2022){Militzer}, {Hubbard}, {Wahl}, {Lunine}, {Galanti}, {Kaspi}, {Miguel}, {Guillot}, {Moore}, {Parisi}, {Connerney}, {Helled}, {Cao}, {Mankovich}, {Stevenson}, {Park}, {Wong}, {Atreya}, {Anderson}, \& {Bolton}}]{militzer2022}
{Militzer}, B., {Hubbard}, W.~B., {Wahl}, S., {et~al.} 2022, The Planetary Science Journal, 3, 185

\bibitem[{{Mirouh} {et~al.}(2012){Mirouh}, {Garaud}, {Stellmach}, {Traxler}, \& {Wood}}]{mirouh2012}
{Mirouh}, G.~M., {Garaud}, P., {Stellmach}, S., {Traxler}, A.~L., \& {Wood}, T.~S. 2012, \apj, 750, 61

\bibitem[{{Moore} {et~al.}(2022){Moore}, {Barik}, {Stanley}, {Stevenson}, {Nettelmann}, {Helled}, {Guillot}, {Militzer}, \& {Bolton}}]{moore2022}
{Moore}, K.~M., {Barik}, A., {Stanley}, S., {et~al.} 2022, Journal of Geophysical Research (Planets), 127, e2022JE007479

\bibitem[{Morales {et~al.}(2013)Morales, Hamel, Caspersen, \& Schwegler}]{morales2013}
Morales, M.~A., Hamel, S., Caspersen, K., \& Schwegler, E. 2013, Phys. Rev. B, 87, 174105

\bibitem[{{Morales} {et~al.}(2009){Morales}, {Schwegler}, {Ceperley}, {Pierleoni}, {Hamel}, \& {Caspersen}}]{morales2009}
{Morales}, M.~A., {Schwegler}, E., {Ceperley}, D., {et~al.} 2009, Proceedings of the National Academy of Science, 106, 1324

\bibitem[{{Mousis} {et~al.}(2016){Mousis}, {Atkinson}, {Spilker}, {Venkatapathy}, {Poncy}, {Frampton}, {Coustenis}, {Reh}, {Lebreton}, {Fletcher}, {Hueso}, {Amato}, {Colaprete}, {Ferri}, {Stam}, {Wurz}, {Atreya}, {Aslam}, {Banfield}, {Calcutt}, {Fischer}, {Holland}, {Keller}, {Kessler}, {Leese}, {Levacher}, {Morse}, {Mu{\~n}oz}, {Renard}, {Sheridan}, {Schmider}, {Snik}, {Waite}, {Bird}, {Cavali{\'e}}, {Deleuil}, {Fortney}, {Gautier}, {Guillot}, {Lunine}, {Marty}, {Nixon}, {Orton}, \& {S{\'a}nchez-Lavega}}]{mousis2016}
{Mousis}, O., {Atkinson}, D.~H., {Spilker}, T., {et~al.} 2016, \planss, 130, 80

\bibitem[{{M{\"u}ller} \& {Helled}(2023)}]{muller2023}
{M{\"u}ller}, S. \& {Helled}, R. 2023, Frontiers in Astronomy and Space Sciences, 10, 1179000

\bibitem[{{M{\"u}ller} {et~al.}(2020){M{\"u}ller}, {Helled}, \& {Cumming}}]{muller2020}
{M{\"u}ller}, S., {Helled}, R., \& {Cumming}, A. 2020, \aap, 638, A121

\bibitem[{{Nettelmann} {et~al.}(2024){Nettelmann}, {Cano Amoros}, {Tosi}, {Fortney}, \& {Helled}}]{nettelmann2024}
{Nettelmann}, N., {Cano Amoros}, M., {Tosi}, N., {Fortney}, J.~J., \& {Helled}, R. 2024, arXiv e-prints, arXiv:2406.16024

\bibitem[{{Nettelmann} {et~al.}(2015){Nettelmann}, {Fortney}, {Moore}, \& {Mankovich}}]{nettelmann2015}
{Nettelmann}, N., {Fortney}, J.~J., {Moore}, K., \& {Mankovich}, C. 2015, \mnras, 447, 3422

\bibitem[{{Pfaffenzeller} {et~al.}(1995){Pfaffenzeller}, {Hohl}, \& {Ballone}}]{pfaffenzeller1995}
{Pfaffenzeller}, O., {Hohl}, D., \& {Ballone}, P. 1995, \prl, 74, 2599

\bibitem[{{Pollack} {et~al.}(1977){Pollack}, {Grossman}, {Moore}, \& {Graboske}}]{pollack1977}
{Pollack}, J.~B., {Grossman}, A.~S., {Moore}, R., \& {Graboske}, H.~C., J. 1977, \icarus, 30, 111

\bibitem[{{Preising} {et~al.}(2023){Preising}, {French}, {Mankovich}, {Soubiran}, \& {Redmer}}]{preising2023}
{Preising}, M., {French}, M., {Mankovich}, C., {Soubiran}, F., \& {Redmer}, R. 2023, \apjs, 269, 47

\bibitem[{{P{\"u}stow} {et~al.}(2016){P{\"u}stow}, {Nettelmann}, {Lorenzen}, \& {Redmer}}]{pustow2016}
{P{\"u}stow}, R., {Nettelmann}, N., {Lorenzen}, W., \& {Redmer}, R. 2016, \icarus, 267, 323

\bibitem[{{Rosenblum} {et~al.}(2011){Rosenblum}, {Garaud}, {Traxler}, \& {Stellmach}}]{rosenblum2011}
{Rosenblum}, E., {Garaud}, P., {Traxler}, A., \& {Stellmach}, S. 2011, \apj, 731, 66

\bibitem[{{Salpeter}(1973)}]{salpeter1973}
{Salpeter}, E.~E. 1973, \apjl, 181, L83

\bibitem[{{Saumon} {et~al.}(1995){Saumon}, {Chabrier}, \& {van Horn}}]{saumon1995}
{Saumon}, D., {Chabrier}, G., \& {van Horn}, H.~M. 1995, \apjs, 99, 713

\bibitem[{{Saumon} {et~al.}(1992){Saumon}, {Hubbard}, {Chabrier}, \& {van Horn}}]{saumon1992}
{Saumon}, D., {Hubbard}, W.~B., {Chabrier}, G., \& {van Horn}, H.~M. 1992, \apj, 391, 827

\bibitem[{{Sch{\"o}ttler} \& {Redmer}(2018)}]{schottler2018_prl}
{Sch{\"o}ttler}, M. \& {Redmer}, R. 2018, \prl, 120, 115703

\bibitem[{{Seiff} {et~al.}(1998){Seiff}, {Kirk}, {Knight}, {Young}, {Mihalov}, {Young}, {Milos}, {Schubert}, {Blanchard}, \& {Atkinson}}]{seiff1998}
{Seiff}, A., {Kirk}, D.~B., {Knight}, T. C.~D., {et~al.} 1998, \jgr, 103, 22857

\bibitem[{{Smoluchowski}(1967)}]{smoluchowski1967}
{Smoluchowski}, R. 1967, \nat, 215, 691

\bibitem[{{Stevenson}(1975)}]{stevenson1975}
{Stevenson}, D.~J. 1975, \prb, 12, 3999

\bibitem[{{Stevenson}(1985)}]{stevenson1985}
{Stevenson}, D.~J. 1985, \icarus, 62, 4

\bibitem[{{Stevenson}(2020)}]{stevenson2020}
{Stevenson}, D.~J. 2020, Annual Review of Earth and Planetary Sciences, 48, 465

\bibitem[{{Stevenson} \& {Salpeter}(1977{\natexlab{a}})}]{stevenson1977b}
{Stevenson}, D.~J. \& {Salpeter}, E.~E. 1977{\natexlab{a}}, \apjs, 35, 239

\bibitem[{{Stevenson} \& {Salpeter}(1977{\natexlab{b}})}]{stevenson1977a}
{Stevenson}, D.~J. \& {Salpeter}, E.~E. 1977{\natexlab{b}}, \apjs, 35, 221

\bibitem[{{Sur} {et~al.}(2024){Sur}, {Su}, {Tejada Arevalo}, {Chen}, \& {Burrows}}]{sur2024}
{Sur}, A., {Su}, Y., {Tejada Arevalo}, R., {Chen}, Y.-X., \& {Burrows}, A. 2024, arXiv e-prints, arXiv:2404.14483

\bibitem[{{Vazan} {et~al.}(2018){Vazan}, {Helled}, \& {Guillot}}]{vazan2018}
{Vazan}, A., {Helled}, R., \& {Guillot}, T. 2018, \aap, 610, L14

\bibitem[{{Vazan} {et~al.}(2016){Vazan}, {Helled}, {Podolak}, \& {Kovetz}}]{vazan2016}
{Vazan}, A., {Helled}, R., {Podolak}, M., \& {Kovetz}, A. 2016, \apj, 829, 118

\bibitem[{{von Zahn} {et~al.}(1998){von Zahn}, {Hunten}, \& {Lehmacher}}]{vonzahn1998}
{von Zahn}, U., {Hunten}, D.~M., \& {Lehmacher}, G. 1998, \jgr, 103, 22815

\bibitem[{{Vorberger} {et~al.}(2007){Vorberger}, {Tamblyn}, {Militzer}, \& {Bonev}}]{vorberger2007}
{Vorberger}, J., {Tamblyn}, I., {Militzer}, B., \& {Bonev}, S.~A. 2007, \prb, 75, 024206

\bibitem[{{Wahl} {et~al.}(2017){Wahl}, {Hubbard}, {Militzer}, {Guillot}, {Miguel}, {Movshovitz}, {Kaspi}, {Helled}, {Reese}, {Galanti}, {Levin}, {Connerney}, \& {Bolton}}]{wahl2017}
{Wahl}, S.~M., {Hubbard}, W.~B., {Militzer}, B., {et~al.} 2017, \grl, 44, 4649

\bibitem[{{Wilson} \& {Militzer}(2010)}]{wilson2010}
{Wilson}, H.~F. \& {Militzer}, B. 2010, \prl, 104, 121101

\bibitem[{{Wood} {et~al.}(2013){Wood}, {Garaud}, \& {Stellmach}}]{wood2013}
{Wood}, T.~S., {Garaud}, P., \& {Stellmach}, S. 2013, \apj, 768, 157

\end{thebibliography}
%
% - join the .bib files when you upload your source files
%-------------------------------------------------------------------

%\onecolumn
\begin{appendix} %First appendix
\section{Comparison of evolution calculations with different EOSs}
\label{appendix:eos}

\begin{figure*}
\centering
   \includegraphics[width=17cm]{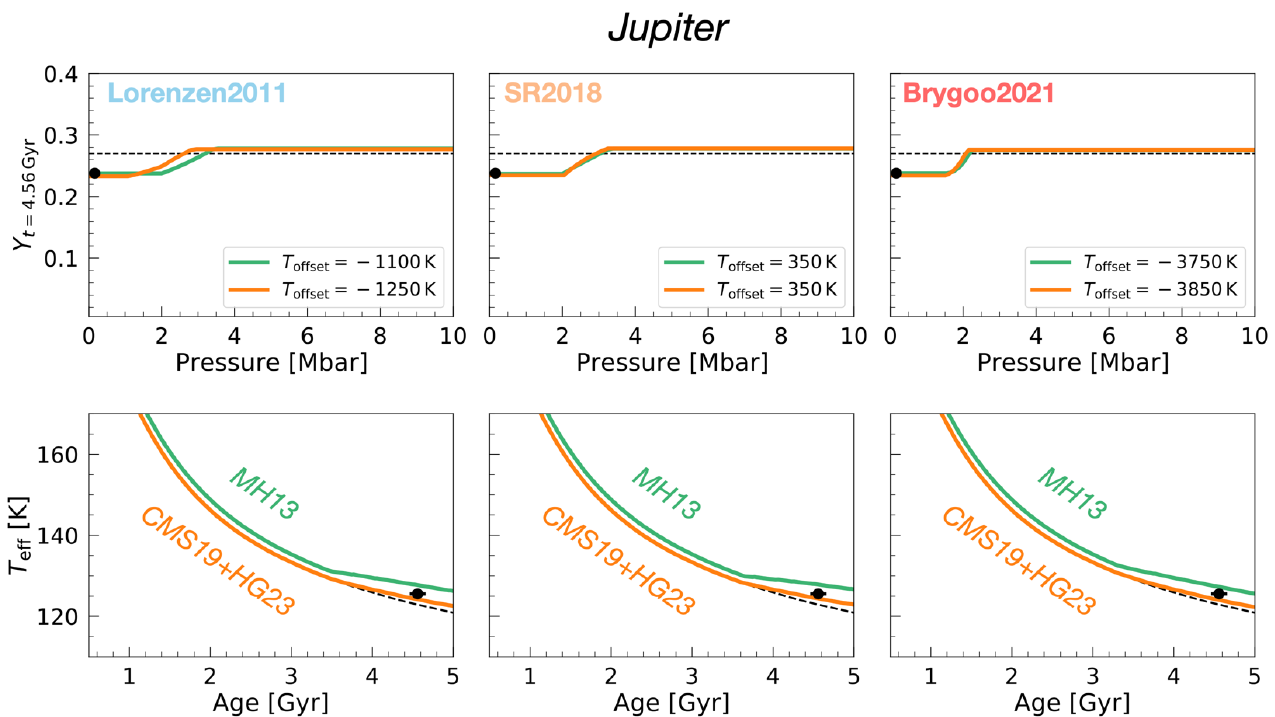}
     \caption{Evolutionary calculations of Jupiter using the CMS19+HG23 and MH13 EOSs. Description is same as for Fig.~\ref{figure:Jupiter_HG23}.}
     \label{figure:Jupiter_MH13}
\end{figure*}

\begin{figure*}
\centering
   \includegraphics[width=17cm]{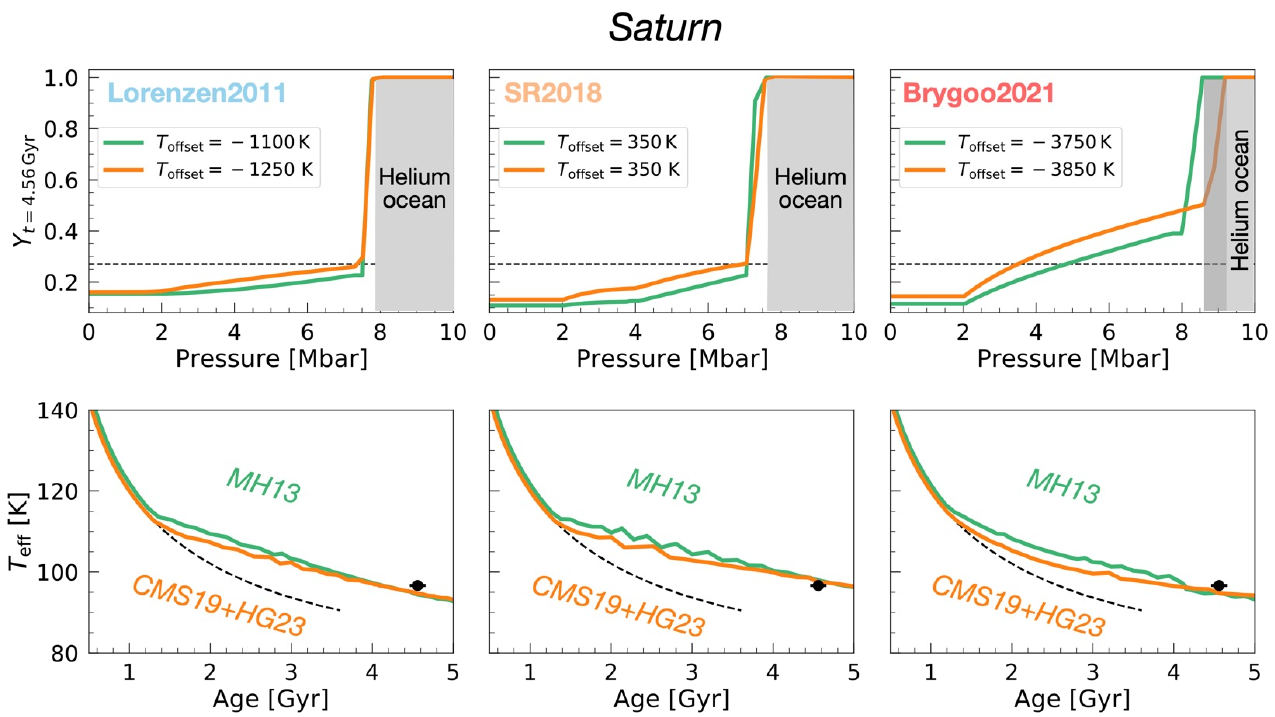}
     \caption{Evolutionary calculations of Saturn using the CMS19+HG23 and MH13 EOSs. Description is same as for Fig.~\ref{figure:Saturn_HG23}.}
     \label{figure:Saturn_MH13}
\end{figure*}

We compare in Figs.~\ref{figure:Jupiter_MH13} and~\ref{figure:Saturn_MH13} our results with the MH13 \citep{militzer2013} and CMS19+HG23 \citep{chabrier2019,howard2023} EOSs for the evolution of Jupiter and Saturn, respectively.

\clearpage
\newpage
\section{Constraining evolution with 1 bar temperature}
%\section{Temperature at 1 bar of baseline models}
\label{appendix:t1bar}

{In this work, we used time to constrain our evolution models, seeking solutions that match the measured atmospheric helium abundance at $t=4.56 \pm 0.1~$Gyr. This time-based approach has been employed in previous studies \citep{fortney2003,mankovich2016,pustow2016,mankovich2020}. However, the 1 bar temperature $T_{\rm 1bar}$, which serves as an outer boundary condition and defines the entropy in the planetary envelope, can also be used to constrain evolution \citep{nettelmann2015,nettelmann2024}. Indeed, as the planet cools, $T_{\rm 1bar}$ decreases and one can track the planetary evolution by looking at the decrease in $T_{\rm 1bar}$. }

{First, we assess whether our baseline models (see Sect.~\ref{subsec:fiducial}) agree with the measured 1 bar temperatures. For Jupiter, Galileo measured $T_{\rm 1bar}=166.1 \pm 0.8~$K \citep{seiff1998}. However, this was measured in a hot spot; the recent reanalysis of the Voyager data yielded $167.3 \pm 3.8~$K and $170.3 \pm 3.8~$K at 0°N and 12°S respectively \citep{gupta2022}. For Jupiter, our baseline models
predict values of the 1 bar temperature of 163.0, 163.0 and 162.6~K for the shifted Lorenzen2011, SR2018 and Brygoo2021 phase diagrams respectively. Our baseline models of Jupiter are therefore in line with the lower bound of the measured $T_{\rm 1bar}$. For Saturn, Voyager measured $T_{\rm 1bar}=135 \pm 5~$K \citep{lindal1985}. We find values of 134.5, 141.0 and 135.2~K for the three shifted phase diagrams respectively, in agreement with the measurement and its upper bound.}

\begin{figure}[h]
   \centering   \includegraphics[width=\hsize]{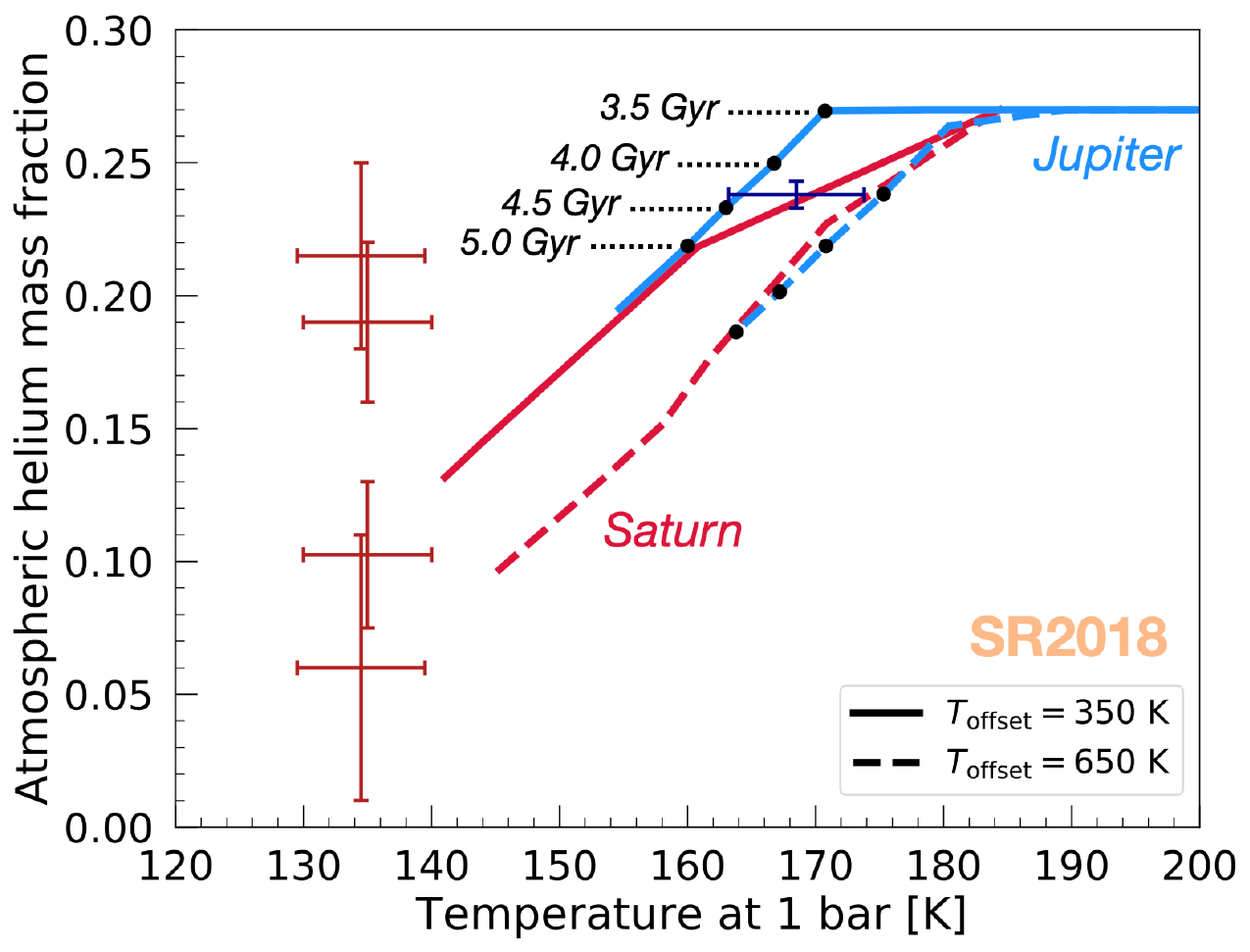}
      \caption{{Inferred atmospheric helium mass fraction as a function of the 1 bar temperature. Solid lines correspond to our baseline models for Jupiter and Saturn, using SR2018 with $T_{\rm offset}=350~K$. Dashed lines show similar evolution models but with $T_{\rm offset}=650~$K. The blue errorbar corresponds to the Galileo measurement of $Y_{\rm atm}$ \citep{vonzahn1998} and the Voyager re-estimation of $T_{\rm 1bar}$ \citep{gupta2022}. Red errorbars (from top to bottom) correspond to estimates from \citet{conrath2000},\citet{koskinen2018},\citet{achterberg2020}, \citet{conrath1984} and the 1 bar temperature from \citet{lindal1985}. Black dots indicate specific ages in the evolution.}}
         \label{figure:t1bar}
\end{figure}

{We next investigate how the 1 bar temperature constraint affects our conclusions. Figure~\ref{figure:t1bar} shows the inferred atmospheric helium mass fraction as a function of $T_{\rm 1bar}$. Our baseline model using SR2018 with $T_{\rm offset}=350~$K yielded for Jupiter $T_{\rm 1bar}=163.0~$K (at $t=4.5~$Gyr), in line with the lower bound of the measured 1 bar temperature. Applying an offset of $T_{\rm offset}=650~$K allows the evolution model to match the upper bound of the measurement (blue dashed line). Using $T_{\rm 1bar}$ rather than the time-based constraint shows that our estimates of the temperature offsets can vary by up to about 300~K. We recall that the uncertainty on the measured atmospheric helium abundance affects those offsets by only $\pm 50~$K. Applying the phase diagram with $T_{\rm offset}$ larger by 300~K to Saturn's evolution could yield an inferred atmospheric helium abundance different by 0.05.}

{Similarly, the planetary radius could also be used to constrain evolution. However, the radius does not evolve much during the last billion years (see Fig.~\ref{figure:core_mass}). Also, since the inferred planetary radius mainly depends on the heavy-element content, matching the radius would only need to adjust the core mass in our models. As shown in Figs.~\ref{figure:offsets_R_T1bar} and~\ref{figure:core_mass}, we find the same values of $T_{\rm offset}$ to fit the measured atmospheric helium abundance, when using a core mass of 10 or 30~$M_{\oplus}$ in Jupiter. Therefore, a small radius mismatch in our baseline models does not affect our conclusions.}

%{We show on Fig.~\ref{figure:appendix_t1bar} the inferred atmospheric helium mass fraction as a function of the 1 bar temperature, for our baseline models.}

%\begin{figure}[h]
%   \centering
%   \includegraphics[width=0.9\hsize]{Images_v4/extrafigure_marked.pdf}
%      \caption{{Inferred atmospheric helium mass fraction as a function of the 1 bar temperature, for our baseline models using the three shifted phase diagrams. Details of the figure are similar to Fig.~\ref{figure:t1bar}.}}
%         \label{figure:appendix_t1bar}
%\end{figure}

\end{appendix}
%---------

\end{document}